\documentclass[12pt,a4paper]{article}
\usepackage[T1]{fontenc}
\usepackage[utf8]{inputenc}
\usepackage{authblk}
\usepackage{bbding} 

\usepackage{amsmath}
\usepackage{graphicx,float,color}
\usepackage{hyperref}
\hypersetup{colorlinks = true, linkcolor = blue, urlcolor  = blue, citecolor = blue, anchorcolor = blue}
\usepackage{multirow}
\usepackage[numbers,sort&compress]{natbib}
\usepackage{braket}
\usepackage{amsfonts}
\usepackage{amssymb} 
\usepackage{epic}
\usepackage{eepic}
\usepackage{epsfig}
\usepackage{latexsym}
\usepackage{color}
\usepackage{url}  
\usepackage{caption}
\usepackage[caption=false]{subfig}
\usepackage{float}
\usepackage{bm}
\usepackage{geometry}
\usepackage{slashed}
\usepackage[detect-all]{siunitx}
\def\Mp{m_{\mathrm{Pl}}}
\def\lp{\ell_{\mathrm{Pl}}}

\def\Mmin{M_{\mathrm{min}}}
\def\hMmin{\hat{M}_{\mathrm{min}}}
\def\Mini{M_{\mathrm{init}}}
\def\hMini{\hat{M}_{\mathrm{init}}}

\def\rm{\mathrm}

\usepackage{lipsum}
\makeatletter
\def\ps@pprintTitle{%
 \let\@oddhead\@empty
 \let\@evenhead\@empty
 \def\@oddfoot{}%
 \let\@evenfoot\@oddfoot}
\makeatother
\usepackage{etoolbox}

\begin{document}
\title{\textbf{Axion dark matter with not quite black hole domination
}}    
\author[]{Ufuk Aydemir\thanks{uaydemir@metu.edu.tr} }
\affil[]{\small Department of Physics, Middle East Technical University, Ankara 06800, T\"urkiye}

\maketitle

\vspace{-0.6cm}
\begin{abstract}
We investigate the effects of an early cosmological period dominated by primordial 2-2-holes on axion dark matter. The 2-2-holes emerge as a new family
of horizonless classical solutions for ultracompact matter distributions in quadratic gravity, a candidate theory of quantum gravity. Thermal 2-2-holes, sourced by relativistic thermal gas, exhibit Hawking-like radiation and fulfill the entropy-area law before they become remnants with almost no radiation. In this paper, we consider the remnant contribution to dark matter (DM) small and adopt the axion DM scenario by the misalignment mechanism.  We show that a 2-2-hole domination phase in the evolution of the universe changes the axion mass window, obtained from the dark matter abundance constraints. The biggest effect occurs when the remnants have the Planck mass, which is the case for a strongly coupled quantum gravity. The change in abundance constraints for the Planck mass 2-2-hole remnants amounts to that of the primordial black hole (PBH) counterpart. Therefore, since we use the revised constraints from gravitational waves on the initial fraction of 2-2-holes, the results here can also be considered the updated version of the PBH case. As a result, the lower limit on the axion mass is found as $m_a \sim10^{-9}$ eV. Furthermore, the domination scenario itself constrains the remnant mass $\Mmin$ considerably. Given that we focus on the pre-BBN domination scenario in order not to interfere with BBN (Big Bang nucleosynthesis) constraints, the remnant mass window becomes $\Mp \lesssim \Mmin \lesssim 0.1\;\rm{g}$.  
\end{abstract}

\newpage
{
  \hypersetup{linkcolor=black}
  \tableofcontents
}

\section{Introduction\label{sec:intro}}

The axion is the (pseudo-) Nambu-Goldstone boson associated with Peccei--Quinn (PQ) $U(1)$ global symmetry and it was originally proposed to explain the strong CP problem of quantum chromodynamics (QCD)~\cite{Peccei:1977hh,Weinberg:1977ma,Wilczek:1977pj}; namely, the question of why the CP violation in the SM from nonperturbative effects turns out to be very small (or absent), which requires the QCD angle $\theta$ constrained to be extremely small (or vanishing). The axion $\phi$ is related to the phase of a complex scalar field charged under the PQ symmetry, which can be generically expressed as $\Phi=\chi\; e^{\phi/f_a}$, where $f_a$ is the symmetry breaking scale of PQ symmetry trough the vacuum expectations of $\Phi$. As a Goldstone boson, the axion is initially massless (at high temperatures) but acquires a small mass through the nonperturbative QCD effects, which introduce explicit breaking terms for the PQ symmetry. Through the axion, the QCD angle is promoted to a field  $\theta =\phi /f_a$, for which $\phi=0$ is energetically favorable. 

Over time, the axion has also emerged as one of the leading DM particle candidates through the misalignment of the minimum of its potential~\cite{Preskill:1982cy,Abbott:1982af,Dine:1982ah}, set by the initial angle $\theta_i$. In order for axion to account for the entire dark matter, it has to be consistent with the relic density constraints. If the axion is the only DM component then the required parameters are $m_a\simeq 10^{-6}$ eV (i.e. $f_a \simeq 10^{12}$ GeV), which otherwise becomes a lower (upper) bound so that the universe is not overclosed (see Refs.~\cite{DiLuzio:2020wdo,Marsh:2015xka} for recent reviews).  

The values for axion parameters given above are valid for standard cosmologies, where the transition from the radiation-dominated era to the matter-domination era in the early-time universe progresses based on the basic model of early cosmology.  On the other hand, in the case of nonstandard cosmologies~\cite{Steinhardt:1983ia,Lazarides:1990xp,Kawasaki:1995vt,Grin:2007yg,Visinelli:2009kt,Nelson:2018via,Allahverdi:2020bys,Heurtier:2021rko,Arias:2021rer,Gomez-Aguilar:2023bej}, which generally include an intermediate stage of early matter domination or kination, the parameter window for the axion sector can change. Recently, Ref.~\cite{Bernal:2021yyb} investigated a nonstandard cosmology triggered by primordial PBH domination and they found that due to the entropy injection of the evaporation of PBHs, the bound from the dark matter abundance on the axion mass becomes $m_a \gtrsim10^{-8}$ eV (or $f_a \lesssim 10^{14}\;\rm{GeV}$) for $\theta_i\sim 1$. 

In this paper, motivated by such effects of PBHs,  we consider a nonstandard cosmology triggered by a period of 2-2-hole domination and investigate the interplay between the 2-2-hole and axion parameter spaces. The 2-2-holes~\cite{Holdom:2002xy, Holdom:2016nek,Holdom:2019ouz,Ren:2019afg}  emanate as horizonless classical solutions for ultracompact distributions in quadratic gravity~\cite{Stelle:1976gc, Voronov:1984kq, Fradkin:1981iu, Avramidi:1985ki}, a candidate quantum theory of gravity (see Refs.~\cite{Salvio:2018crh,Alvarez-Gaume:2015rwa} for a recent review of quadratic gravity).
A 2-2-hole resembles a black hole in the exterior without the event horizon, whereas in the interior, it is characterized by a distinct high curvature solution with a transition region around the would-be horizon. These objects, as black-hole mimickers (i.e.~horizonless ultracompact objects~\cite{Cardoso:2019rvt}),  are expected to appear as dark as black holes for a distant observer, despite lacking event horizons, due to light being trapped in the high redshift region in their deep gravitational potential~\cite{Holdom:2016nek,Cardoso:2019rvt}. Even though the observed ultracompact objects appear to be consistent with General Relativity, confirming these objects as black holes requires precise tests of near-horizon (or would-be horizon) physics, and this is an active field of study~\cite{Cardoso:2019rvt}. In particular, binary mergers of such objects are expected to cause distinctive signatures in gravitational wave signals as late-time echoes in the waveforms~\cite{Cardoso:2019rvt}. There have been some discussions regarding the existence of statistically significant signals of such nature in the current data~\cite{Abedi:2016hgu,Westerweck:2017hus,Abedi:2018pst}, where no consensus has been reached. As for 2-2-holes, the characteristics of gravitational wave echoes have been studied in Refs.~\cite{Conklin:2017lwb,Conklin:2019fcs}. Any sign of the existence of these objects will have profound consequences regarding gravity beyond the theory of General Relativity. 

In contrast to many other ultracompact objects, a 2-2-hole can possess an arbitrarily large mass, yet its mass is bounded from below, indicating the existence of remnants. Until the remnant stage of their evaporation,  2-2-holes exhibit Hawking-like radiation and satisfy the entropy-area relation. Remarkably, these properties directly arise from self-gravitating relativistic thermal gas on a curved background, without taking into account spontaneous particle creation from the vacuum, and thus originate from a completely different origin than the black holes.\footnote{Taking into account particle creation effects from the vacuum on the curved background for 2-2-holes should introduce the usual contribution to the temperature and entropy (as in the case of black holes), expected to be on the order of Hawking temperature and  Bekenstein--Hawking entropy, respectively.} Once they become remnants, 2-2-holes behave more like an ordinary thermodynamic system with an extremely slow pace of radiation. Therefore, the remnants can be considered cold and stable objects, and hence constitute a viable dark matter (DM) candidate~\cite{Aydemir:2020xfd}. Furthermore 2-2-holes, just like black holes, can produce particles that are responsible for the baryon asymmetry, dark matter, and dark radiation~\cite{Aydemir:2020pao}.

Axion cosmology, as noted above, is another area of interest that can be connected to 2-2-hole density evolution in the early universe, which is the focus of this paper. Naturally, a 2-2-hole domination period, just in the case of black holes~\cite{Bernal:2021yyb}, can change the evolution of the Hubble rate and, therefore, can affect the axion abundance constraints. This is due to the large entropy injection at the end of the nonremnant stage of 2-2-hole evaporation, during which the radiation is Hawking-like. Once the 2-2-holes become remnant, they behave like cold and stable matter, and therefore contribute to the DM abundance. Since we are interested in the axion DM case, we only focus on parameter space where the remnant contribution to DM is negligible. In addition to the misalignment mechanism, axions can also be produced from the 2-2-hole evaporation process. As could be inferred from our previous study~\cite{Aydemir:2020pao}, this contribution is negligible compared to the DM abundance; moreover, the axions directly produced from 2-2-holes can be hot and can make a meaningful contribution to the effective number of relativistic degrees of freedom in the early universe, $N_{\rm{eff}}$, which will be briefly discussed later in this paper. Due to the extreme constraints from Big Bang nucleosynthesis (BBN)~\cite{Aydemir:2020xfd}, we only focus on 2-2-holes small enough to complete their evaporation by the beginning of the BBN period. This maximum mass, of course, has a dependence on the remnant mass $\Mmin$. 

Our scenario deviates from the PBH case in several aspects. First, we have naturally leftover remnants, whereas it has been so far a conjecture for PBHs if one prefers to adopt the remnant scenario. Note that the 2-2-hole provides a specific example of how high curvature terms can play a role in preventing horizon formation, which is also anticipated for the possible existence of BH remnants.  Due to the dependence on the remnant mass $\Mmin$, the axion abundance limits can constrain this parameter. The pre-BBN 2-2-hole domination scenario itself limits $\Mmin$, where the most strict bound, $\Mmin\lesssim 0.1\;\rm{g}$, comes from the condition that the background temperature when the 2-2-hole domination period begins, $T_{\rm{dom}}$, should be higher than the temperature when 2-2-holes become remnants $(T_{\tau})$. Second, since the radiation rate depends on the remnant mass, in addition to the initial mass of the hole, the required initial mass to survive up to some critical times can be much larger than the PBH counterpart for a given  $\Mmin$. This is because this upper limit on the initial mass goes with $\propto(\Mmin/\Mp)^{2/3}$, where $\Mmin\gtrsim \Mp$. Finally, since the Planck mass 2-2-hole remnant scenario ($\Mmin\simeq3.6\;\Mp$, to be more precise) parametrically amounts to the PBH case; since we are using the revised constraints from the GWs during BBN~\cite{Domenech:2020ssp} (see Refs.~\cite{Papanikolaou:2020qtd,Inomata:2020lmk} for earlier discussions), our study can also be considered the updated version of the PBH case in this limit. As we will show, the lowest bound on the axion mass occurs in the Planck mass limit, which yields $m_a \gtrsim 10^{-9}\;\mathrm{eV}$.

The rest of the paper is organized as follows. In Section~\ref{sec:Axion-review}, we briefly review the standard axion cosmology. In Section~\ref{sec:domination}, we discuss the primordial 2-2-hole domination scenario and the necessary conditions. In Section~\ref{sec:entropy-injection}, the effects of entropy injection from the 2-2-hole evaporation, are investigated.  We present the main results in Section~\ref{sec:results} and conclude with Section~\ref{Conclusion}. 


\section{QCD Axions as Dark Matter in the Standard Cosmology \label{sec:Axion-review}}
The axion dynamics is conjectured as a mechanism to naturally drive the CP-violating term in the SM to zero and hence provides a resolution to the strong CP problem in QCD.  The axion is initially massless as the Nambu-Goldstone boson of the spontaneously broken global $U(1)_{\rm{PQ}}$ symmetry, namely Peccei--Quinn (PQ) symmetry~\cite{Peccei:1977hh,Weinberg:1977ma,Wilczek:1977pj}. Once the nonperturbative effects, which arise due to the axion's couplings to QCD, become important at around the QCD scale, the axion acquires temperature-dependent mass, which was determined by methods in lattice QCD as~\cite{Borsanyi:2016ksw}
\begin{eqnarray}\label{eqn:m(T)}
\tilde{m}_a(T)\simeq m_a \times \begin{cases}
&(T_{\mathrm{QCD}}/T)^4\;\quad \textrm{for} \quad T \gtrsim T_{\mathrm{QCD}} \\
&1\;\qquad\qquad \quad \textrm{ for}\quad T\leqslant T_{\mathrm{QCD}}\;\;,
\end{cases}
\end{eqnarray}
where $T_{\mathrm{QCD}}\simeq150$ MeV.  The relation between the constant mass factor $m_a$ and the decay constant $f_a$ (which is set to be around the Peccei--Quinn symmetry breaking scale) is given as~\cite{GrillidiCortona:2015jxo}
\begin{eqnarray}\label{eqn:axionmass}
\frac{m_a}{\mathrm{eV}}=5.70\times10^6\;\frac{\mathrm{GeV}}{f_a}\;.
\end{eqnarray}

 The Lagrangian for the axion $\phi$, a real scalar field minimally coupled to gravity, in the massive realm  is given simply as
\begin{eqnarray}
\label{Lagrangian}
\mathcal{L}_a=-\frac{1}{2} (\partial _{\mu}\phi) \partial ^{\mu}\phi-\frac{1}{2}m_a^2\phi^2\;.
\end{eqnarray}

From varying the action $\mathcal{S}=\int\sqrt{-g} \;\mathcal{L}_a$ on the Friedmann--Lemaître--Robertson--Walker (FLRW) background, $g_{\mu\nu}=\mathrm{diag}(-1, a^2, a^2, a^2)$ with $a(t)$ being the scale factor,  the equation of motion for the spatially homogeneous axion field becomes
\begin{eqnarray}\label{eqn:eomaxion}
\ddot{\phi}+3 H \dot{\phi}+\tilde{m}_a^2(T) \phi=0\;,
\end{eqnarray}
where $H=\dot a/a$ is the Hubble parameter, as usual. Therefore, the system behaves like a damped harmonic oscillator where the critical damping point occurs for $T_{\mathrm{osc}}$ where $H(T_{\mathrm{osc}}) \sim \tilde{m}_a (T_{\mathrm{osc}})$. In general, this condition is chosen as $3H(T_{\mathrm{osc}}) = \tilde{m}_a (T_{\mathrm{osc}})$. When  $\tilde{m}_a (T)\lesssim 3H(T)$, which corresponds to $T\gtrsim T_{\mathrm{osc}}$, the system is overdamped.
Thus, for very high temperatures ($\tilde{m}_a \approx 0$), the axion field does not really move, and the solution for Eq.~(\ref{eqn:eomaxion}) is $\phi = \mathrm{constant}$. On the other hand, for $T\lesssim T_{\mathrm{osc}}$ (i.e. $\tilde{m}_a (T)\gtrsim 3H(T)$),  the system describes the damped harmonic motion, and the axion field begins to oscillate with angular frequency $\tilde{m}_a (T)$. Therefore, this production of axions is a nonthermal process whose scale is set by the Hubble parameter, which can be related to the corresponding temperature through the Friedmann equation, 
\begin{eqnarray}
 \frac{ H^2(T) M_{\mathrm{Pl}}^2}{8\pi}=\frac{\rho_R(T)}{3}=\frac{\pi^2}{90}g_{*,R} (T) T^4\;,
\end{eqnarray}
for a radiation-dominated epoch of the universe. Here, $M_{\mathrm{Pl}}$ is the Planck mass, and $g_{*,R}(T)$ is the effective number of relativistic degrees of freedom contributing to the radiation energy density $\rho_R(T)$ at temperature $T$. 

In order to find the axion relic density, we will utilize the entropy conservation. In the standard cosmological scenario (which will change later in our discussion), it is generally assumed that no entropy is generated after the axion field begins to oscillate, and therefore, the total entropy $S=s a^3$ is conserved. Since the coherent axion oscillations  behave like nonrelativistic matter, its number density $n_{a}(T)=\rho_a(T)/m_a(T)$ varies with $a^{-3}$, just like the entropy density $s$. Since, $n_a/s$ remains constant as long as there is no entropy injection/production through the evolution of the universe, we can determine the current energy density of axions from
\begin{eqnarray}
\rho_a\left(T_0\right)=\rho_a\left(T_{\mathrm{osc}}\right) \frac{m_a}{\tilde{m}_a\left(T_{\mathrm{osc}}\right)} \frac{s\left(T_0\right)}{s\left(T_{\mathrm{osc}}\right)}\;,
\end{eqnarray}
where $T_0$ is the temperature today and since $T_0\ll T_{\mathrm{QCD}}$, $\tilde{m}_a\left(T_{0}\right)=m_a$ from Eq.~(\ref{eqn:m(T)}). Note that the entropy density is given as
$s(T)=\frac{2 \pi^2}{45} g_{* s}(T) T^3$, whose value at present is $s\left(T_0\right) \simeq 2.69 \times 10^3 \mathrm{~cm}^{-3}$~\cite{Planck:2018vyg}. Note that $g_{* s}(T)$ denotes the number of relativistic degrees of freedom contributing to the entropy. The difference between $g_{* s}(T)$ and $g_{*,R}$ (defined above) is small and generally ignored, i.e. $g_{*,R}\approx g_{* s}(T)$. The energy density of the axion field due to its misalignment from the minimum of its potential is given as $\rho_a\left(T_{\mathrm{osc}}\right) \simeq \frac{1}{2} \tilde{m}_a^2\left(T_{\mathrm{osc}}\right) f_a^2 \theta_i^2$, where $\theta_i$ is the initial misalignment angle. The axion abundance $\Omega_a h^2  \equiv \frac{\rho_a\left(T_0\right)}{\rho_c / h^2}$   is then found as
\begin{eqnarray}
\begin{aligned}
\frac{\Omega_a h^2}{0.12} 
& \simeq \left(\frac{\theta_i}{10^{-3}}\right)^2\times \begin{cases}\left(\dfrac{m_a}{m_a^{\mathrm{QCD}}}\right)^{-\frac{3}{2}} & \text { for } m_a \leq m_a^{\mathrm{QCD}} \\
\left(\dfrac{m_a}{m_a^{\mathrm{QCD}}}\right)^{-\frac{7}{6}} & \text { for } m_a \geq m_a^{\mathrm{QCD}}\;,\end{cases}
\end{aligned}
\end{eqnarray}
where we follow the notation of Ref.~\cite{Bernal:2021yyb}. Here,  $m_a^{\mathrm{QCD}} \equiv m_a\simeq 4.8 \times 10^{-11} \mathrm{eV}$ is the axion mass value for $T_{\mathrm{osc}}=T_{\mathrm{QCD}}$.  $\rho_c / h^2 \simeq 1.1 \times 10^{-5} \;\mathrm{GeV} / \mathrm{cm}^3$ is the critical energy density in terms of Hubble parameter $h$, and  $\Omega_a h^2 \simeq 0.12$ is the required value for the axion abundance to account for the entire DM~\cite{Planck:2018vyg}.

The expected range of values for the initial misalignment angle depends on whether or not the PQ symmetry breaking occurs after a possible period of inflation~\cite{Marsh:2015xka}. If the universe never underwent inflation or the PQ symmetry breaking occurred before inflation ($H_I< f_a$), then the initial misalignment angle $\theta_i$ is a free parameter, and its expected (or "natural") interval is taken as $0.5<\theta_i<\pi / \sqrt{3}$ (where $\pi / \sqrt{3}=(\theta_i)_{\rm{rms}}$). Much smaller values are also possible depending on how much fine-tuning one is willing to allow. Such small values, on the other hand, are sometimes considered in the anthropic perspective in the literature. On the other hand, if the PQ symmetry occurred after inflation ($H_I> f_a$), $\theta_i$ is not a free parameter and it is averaged over the uniform distribution in the interval $[-\pi,\pi]$, namely $\theta_i=\pi/\sqrt{3}$. In the latter case, topological defects can arise due to PQ symmetry breaking and have to be dealt with.  Moreover, in this scenario (i.e. $H_I> f_a$), the backreaction effects become important and the parameter space becomes quite restricted such that the region $f_a \lesssim 10^{15}$ GeV (or $m_a\gtrsim 10^{-9}$ eV) is excluded in order not to produce too much dark matter~\cite{Marsh:2015xka}.

Therefore, the standard axion scenario, in general, assumes that $H_I\ll f_a$, where the PQ symmetry breaking occurred before inflation and $\theta_i$ is a free parameter; this is the case we adopt in this paper as well. The classic axion window is taken as $0.5<\theta_i<\pi / \sqrt{3}$, corresponding to $10^{-6}\; \mathrm{eV}$ $\lesssim m_a \lesssim 10^{-5}\; \mathrm{eV}$ (or $10^{11}\; \mathrm{GeV}$ $\lesssim f_a \lesssim 10^{12}\; \mathrm{GeV}$). We will investigate the effects of primordial 2-2-domination in this parameter space.

Note that the potential term in the Lagrangian is a generic (harmonic) approximation to a more complicated, model-dependent, periodic potential for small displacements ($\theta=\phi/f_a <1$) from the potential minimum~\cite{Marsh:2015xka}. Anharmonic corrections may become important for large initial angle values, $\theta_i\simeq1$. Such corrections can be introduced pragmatically to the solution of the linear equation, given in Eq.~(\ref{eqn:eomaxion}), for the full nonlinear solution by employing the replacement $\theta_i^2\rightarrow \theta_i^2 f_{\mathrm{anh}}(\theta_i)$, where $f_{\mathrm{anh}}(\theta_i)$ is called anharmonicity correction factor. For instance, such a correction factor to cosine potential was found by Ref.~\cite{Visinelli:2009zm} as
\begin{eqnarray}\label{fann}
f_{\mathrm{anh}}(x)=\left[\ln \left(\frac{e}{1-x^2 / \pi^2}\right)\right]^{7 / 6}\;,
\end{eqnarray}
which we will use below when we take into account such effects.

\section{Primordial 2-2-hole domination \label{sec:domination}}     

As we have seen above, we have quite restricted parameter space for the axion to account for all DM in the standard cosmological scenario. Ref.~\cite{Bernal:2021yyb} considered the scenario that there was an era in the evolution of the universe in which PBH came to dominate the total energy density. This introduces an intermediate stage of matter domination that relaxes the axion bounds by a couple of orders of magnitude.

Now, we will look at a similar situation for the 2-2-holes. Just like PBHs, 2-2-holes also behave like cold matter which evolves with $a^{-3}$. But differently, we have a well-established notion of 2-2-holes remnants, whose mass $M_{\mathrm{min}}$ is a free parameter and comes from the underlying theory of quantum gravity (quadratic gravity). (In the case of BHs also, sometimes remnants are conjectured to occur although the mechanism of that is not clear and in general assumed to have Planck mass remnants, unlike in our case.) In fact, we studied the 2-2-hole remnants as whole DM in Ref.~\cite{Aydemir:2020xfd} and as complimentary to particle DM in Ref.~\cite{Aydemir:2020pao}. 
Therefore, in the case of axion DM we need to be careful to take into account the necessary constraints.

To avoid interfering with the BBN constraints, we will focus on the 2-2-holes that become remnants by the beginning of the BBN era ($t\sim1$ s). The mass fraction of 2-2-hole remnants in dark matter today is
\begin{eqnarray}\label{eq:betaf}
r \equiv\frac{\Mmin\,n(t_0)}{\rho_\textrm{DM}(t_0)}
=\frac{\Mmin\,s(t_0)}{\rho_\textrm{DM}(t_0)}\frac{n(t_0)}{s(t_0)}\,,
\end{eqnarray} 
where $n(t)$ denotes the remnant number density, $s(t_0)=2.9\times 10^3\,\textrm{cm}^{-3}$ and $\rho_{\textrm{DM}}(t_0)\approx 0.26\rho_c$ are the entropy and dark matter densities today, with $\rho_c=9.5\times10^{-30}\,\textrm{g}\,\textrm{cm}^{-3}$ being the critical density~\cite{Planck:2018vyg}.\footnote{Another commonly used parameter in the literature is the mass fraction at the formation of the holes, namely $\beta\equiv\rho(t_\textrm{init})/\rho_\textrm{tot}(t_\textrm{init})$~\cite{Carr:2009jm}. For 2-2-holes, it is related to the remnant fraction $r$ as $\beta\approx 4.0\times 10^{-28} \,f\,\hMmin ^{-1} \,\hMini^{3/2}$, where the background temperature $T_\textrm{bkg}(t)=0.17 \,\Mp\, (t/\lp)^{-1/2}$ and the inital 2-2-hole mass $\Mini\approx 8\times 10^{37}\left(t_\textrm{init}/\textrm{s}\right)$\,g~\cite{Aydemir:2020xfd} is used.} Note that here, as the leading order approximation for the cosmic evolution, we consider the evaporation as instantaneous radiation of energy at $t\approx \tau_L$\footnote {This is commonly called as $t_{\rm{eva}}$ for the black hole case in the literature. For us, it is the end of the large-mass stage (away from the minimum mass $\Mmin$) of the 2-2-hole radiation, where the radiation occurs in a Hawking-like behavior. After  $\tau_L$, the 2-2-holes radiate like coal with an extremely slow rate and effectively become cold remnants.}, with the 2-2-hole mass $M(t) \approx \Mini$ at $t \leq \tau_L$ and $M(t) \approx \Mmin$ at $t > \tau_L$.  Note that the background temperature in the radiation-dominated early universe is given as
\begin{eqnarray}\label{bkgtemp}
T(t)=1.6\; g_{*}^{-1/4}\;(t/\rm{s})^{-1/2}\; \rm{MeV}\;,
\end{eqnarray}
which should not be confused with the temperature of the 2-2-hole radiation, $T_{\infty}(t)$.

The relation between $r$ and the number density to entropy density ratio at the time of formation, $n(t_\textrm{init})/s(t_\textrm{init})$, depends on whether or not the primordial 2-2-holes ever came to dominate the energy density. Even though the initial mass fraction of 2-2-holes starts with a small value (in the radiation era of the very early universe), since they behave as cold nonrelativistic matter and therefore evolve with $a^{-3}$ as opposed to the radiation which goes with $a^{-4}$, their relative mass fraction increases with time. If we define the ratio of 2-2-hole energy density to the radiation energy density as $\beta\equiv\rho (t_{\mathrm{init}})/\rho_R (t_{\mathrm{init}})= M _{\mathrm{init}} n(t_\textrm{init})/\rho_R (t_{\mathrm{init}})$, the condition that the 2-2-holes and the radiation have equal energy densities before 2-2-holes become remnant ($t\approx \tau_L$), i.e. the condition for the 2-2-hole domination to occur, can be given in terms of a critical value $\beta_c\simeq T(\tau_L)/T(t_{\mathrm{init}})$, which can be expressed in terms of the critical number density as~\cite{Aydemir:2020xfd}, 
\begin{eqnarray}\label{critical}
 \beta_c &\simeq&  M _{\mathrm{init}} n_c(t_\textrm{init})/\rho_R (t_{\mathrm{init}}) \quad\quad\mathrm{where} \quad\quad n_c(t_\textrm{init})\simeq\frac{\rho_\textrm{rad}(t_\textrm{init})}{\Mini}\sqrt{\frac{t_\textrm{init}}{\tau_L}}\,\;,\nonumber\\
 &\simeq& 3.5\times10^{-2}\; \frac{\Mmin}{\Mini}\;.
 \end{eqnarray}

If $n(t_\textrm{init})\lesssim n_c(t_\textrm{init})$  (i.e. $\beta\lesssim \beta_c$),  the 2-2-holes are always subdominant in the energy budget, and the entropy injection from evaporation is negligible. The ratio $n(t)/s(t)$ remains constant till the present, with $n(t_0)/s(t_0) \approx n(\tau_L)/s(\tau_L)\approx n(t_\textrm{init})/s(t_\textrm{init})$. The mass fraction of remnants today is then $r\approx 2.6\times 10^{28}\hMmin \;n(t_\textrm{init})/s(t_\textrm{init})$. This is the nondomination scenario, and we are not interested in this case since it does not affect the evolution of the universe significantly.

If, on the other hand, $n(t_\textrm{init})\gtrsim n_c(t_\textrm{init})$ (i.e. $\beta\gtrsim \beta_c$), then the domination scenario occurs, in which there is a new era of matter domination before $\tau_L$. It turns out that the extra redshift of the number density introduced by this new era cancels with the large initial density so that $n(\tau_L)$ remains the same as the one with $n_c(t_\textrm{init})$, i.e. $n(t_\textrm{init})=n_c(t_\textrm{init})$~\cite{Aydemir:2020xfd}. 
Thus, the mass fraction at present has a maximum, 
\begin{eqnarray}
\label{fmax}
r_{\mathrm{max}} 
= 2.6\times 10^{28}\hMmin \frac{n_c(t_\textrm{init})}{s(t_\textrm{init})}
= 9.1\times 10^{25}\, \hMmin ^{2} \,\hMini^{-5/2}\,,
\end{eqnarray}
and the bound is saturated with $r =r_\textrm{max}$ for the domination scenario. The "hat" notation denotes the Planck mass normalized mass values, e.g., $\hMmin\equiv\Mmin/\Mp$, where $\Mp=2.2\times 10^{-5}\;\rm{g}$ is the Planck mass. 
There is a critical value of 2-2-hole initial mass, say $M_\textrm{c}$, corresponding to $r_\textrm{max}=1$, given as 
\begin{eqnarray}
\label{Mc}
M_\textrm{c}=
5.3\times 10^5\,
\hMmin^{4/5}\;\textrm{g}.
\end{eqnarray}
Thus, for $\Mini\lesssim M_\textrm{c}$, with $r_{\textrm{max}}$ being greater than unity, the 2-2-hole remnants can account for all of the dark matter since $r\approx 1$ is allowed, as we investigated in Ref.~\cite{Aydemir:2020xfd}. In this case, the 2-2-hole domination is not allowed due to the fact that this would require $r\approx r_{\textrm{max}}$ (the domination condition) and thus $r >1$, which is forbidden.  As for $\Mini\gtrsim M_\textrm{c}$ on the other hand, the 2-2-hole domination can occur since $r_{\textrm{max}}\lesssim 1$ and hence the domination condition ($r\approx r _{\textrm{max}}$) is allowed, and in this case the remnants cannot account for whole dark matter. The latter scenario is the one we are interested in here since we want 2-2-hole domination in the early universe, and we want 2-2-hole remnants not being the majority of DM since that role is saved for axions. For concreteness, we choose the value $r_{\textrm{max}}\lesssim 0.01$ in order to guarantee a small 2-2-hole abundance. Therefore, from Eq.~(\ref{fmax}) we determine a second critical initial mass for 2-2-holes as
\begin{eqnarray}
\label{Mc2}
M_{\bar{\textrm{c}}}=
3.3\times 10^6\,
\hMmin^{4/5}\;\textrm{g}.
\end{eqnarray}
Together with the aforementioned condition that the 2-2-holes of interest become remnants by the beginning of BBN, the initial mass of 2-2-holes must be in the interval
\begin{eqnarray}
\label{massinterval}
M_{\bar{\textrm{c}}}\lesssim \Mini\lesssim M_\textrm{BBN}
\end{eqnarray}
where $M_\textrm{BBN}$ is the initial mass of a 2-2-hole that would evaporate (and become remnant) at the beginning of BBN ($T_{\rm{BBN}}\gtrsim 4$ MeV). To find $M_\textrm{BBN}$, if we assume the noninterrupted radiation domination as in the case of standard cosmology, we can use Eq.~(\ref{bkgtemp}) to find that $t_{\rm{BBN}}\lesssim 10^{-2}\;\rm{s} $. If we take the condition $\tau_L\leqslant t_{\rm{BBN}}$, we get a value for the critical initial mass for the 2-2-holes. But we do not adopt the standard cosmology in our scenario, since we consider a 2-2-hole (matter) domination era before BBN. Once the 2-2-holes complete their large-mass stage evaporation (and become remnants), they reheat the universe to a temperature $T_{\tau}$, and the universe becomes radiation-dominated again. Now, instead of using Eq.~(\ref{bkgtemp}), we can find $T_{\tau}$ from the Friedmann equation for the radiation-dominated new phase, as given in Eq.~(\ref{reheatingtemp}).  By imposing $T_{\tau}\gtrsim T_{\rm{BBN}}$, we find the value for $M_\textrm{BBN}$ as
\begin{eqnarray}\label{MBBN}
M_\textrm{BBN}= 1.4\times 10^8 \,\hMmin^{2/3}\;\textrm{g} \,.\quad
\end{eqnarray} 
which interestingly does not differ noticeably from the value that can be found in the standard cosmological scenario, as explained above.

For the later discussion, an important input is the 2-2-hole number density to entropy ratio right after evaporation. For the domination case, i.e. $n(t_\textrm{init})=n_c(t_\textrm{init})$, we have~\cite{Aydemir:2020xfd}
\begin{eqnarray}
\label{novers}
\frac{n(\tau_L)}{s(\tau_L)}=
 3.5\times10^{-3}\, \hMmin\, \hMini^{-5/2}\,.
\end{eqnarray}
As mentioned above, this is the result of the entropy injection due to the 2-2-hole evaporation following a 2-2-hole dominated era in the evolution of the universe. 

 \subsection{The background temperature of the universe at critical times}
 Recall that we can divide the evolution of the universe (before BBN) into four regimes: $t\leqslant t_{\mathrm{eq}}$, 
$t_{\mathrm{eq}}< t \leqslant t_{\mathrm{dom}}$, $t_{\mathrm{dom}}< t \leqslant \tau_L$, and $\tau_L <t$. For each regime, the Hubble parameter can be found in terms of background temperature, as we will do in the next subsection. But first, let us find the background temperatures corresponding to $t=t_{\mathrm{eq}}$ and $t=\tau_L$.  

At $t\simeq \tau_L$, where 2-2-holes complete its large-mass stage evaporation (and become remnants), the universe is now filled with radiation and becomes radiation-dominated again. Therefore, we can find the background temperature at $t\simeq \tau_L$, i.e. $T (\tau_L)\equiv T_\tau$, from the Friedmann equation\footnote{Notice that we have the identical  Friedmann equation to the GR case. This is mainly because we consider the $R^2$ term in the quadratic Lagrangian (Eq.~(\ref{action})) negligible since this term does not play an important role for 2-2-hole solutions, and it is ignored for simplicity reasons while finding the particular solution we use in this paper~\cite{Ren:2019afg}. Note also that there is a formulation of quadratic gravity that removes the $R^2$ term by introducing an auxiliary scalar, followed by some field redefinitions~\cite{Salvio:2018crh}).~Meanwhile, the Weyl squared term in the action vanishes for the FRW metric and does not play a cosmological role. This is an advantage for the purpose of this paper since we would like to explore the axion DM parameter space with modifications in the standard cosmological scenario brought by only the intermediate 2-2-hole domination era before BBN while not altering the rest of the cosmological evolution.}
\begin{eqnarray}
\label{RHtemp}
\frac{3 \Mp^2 H^2 (\tau_L)}{8\pi}\simeq\rho_{R}(\tau_L)=\frac{\pi^2}{30}g_{*}(T_{\tau})T^4_{\tau}\;.
\end{eqnarray}
Since the universe was matter-dominated from $t=t_{\rm{d}}$, we can use the Hubble parameter $H\simeq 2/(3\tau_L)$. By using Eqs.~(\ref{RHtemp}) and (\ref{timevalues}), we find 
\begin{eqnarray}
\label{reheatingtemp}
T_\tau = 6.9 \times 10^{12}\; \hat{M}_{\mathrm{min}} \left(\frac{{M}_{\mathrm{init}}}{\rm{g}}\right) ^{-3 / 2}\;\mathrm{MeV}\;,
\end{eqnarray}
where $\hat{M}_{\mathrm{min}}\equiv M_{\mathrm{min}}/\Mp$, as before. Note that, as in the case of BHs, this value can also be called reheating temperature since the universe reheated to this value after the 2-2-hole evaporation, which, of course, has nothing to do with the reheating after inflation. As we discussed before, in order to avoid disrupting BBN, we require that $T_\tau\gtrsim 4\;\rm{MeV}$, which amounts to imposing $\Mini\leqslant M_{\rm{BBN}}$, whose value is given in Eq.~(\ref{MBBN}).  As explained around Eq.~(\ref{massinterval})), we also have a lower bound for $\Mini$ as a result of the domination condition. Therefore, the condition  $M_\textrm{c}\lesssim \Mini\lesssim M_\textrm{BBN}$ yields that the reheating temperature should satisfy
\begin{eqnarray}
\label{reheatinginterval}
4\;\rm{MeV}\lesssim T_\tau \lesssim 1.2\times 10^3\; \hMmin^{-1/5}\;\rm{MeV}\;.
\end{eqnarray}
Since the minimum value for $\Mmin$ is the Planck mass $\Mp=2.18\times 10^{-5}\;\rm{g}$, the upper limit for the reheating temperature is $\sim 1\;\rm{GeV}$. Notice that from Eq.~(\ref{reheatinginterval}) we can also determine the upper limit for the remnant mass $\Mmin$ (see Eq.~(\ref{Mmininterval})). 
  
From the time the 2-2-holes formed ($t=t_{\rm{init}}\approx (8\times 10^{37})^{-1} \Mini\;\rm{s} /\rm{g} $) up to the time $t=t_{\mathrm{eq}}$, the 2-2-holes are subdominant, and therefore there is no significant entropy injection during this time.
 By using the entropy conservation, we can find the background temperature of the universe at $t=t_{\mathrm{eq}}$ as
\begin{eqnarray}
\label{Teq}
T_{\mathrm{eq}}=\beta \; T_{\mathrm{init}}\left(\frac{g_{* s}\left(T_{\mathrm{init}}\right)}{g_{* s}\left(T_{\mathrm{eq}}\right)}\right)^{1 / 3}\;,\quad\rm{where}\quad T_{\mathrm{init}}= T(t_\mathrm{init})=4.3\times 10^{15}\left(\frac{\Mini}{\rm{gr}}\right)^{-1/2}\rm{GeV}\;,
\end{eqnarray}
where we recall that $g_{* s}\simeq g_{*}$ at high energies~\cite{Husdal:2016haj} and where we use Eq.~(\ref{bkgtemp}) to obtain the background temperature at the time of 2-2-hole formation, $T_{\mathrm{init}}$. Notice that $\mathrm{eq}$ in  $T_{\mathrm{eq}} (t_{\mathrm{eq}}$) does not denote the usual matter-radiation equality, which is right around CMB and is denoted in this paper with the capital letters as in  $T_{\mathrm{EQ}} (t_{\mathrm{EQ}})$. We also take $\beta \gtrsim \beta_c$ (or $n(t_\textrm{init})\gtrsim n_c(t_\textrm{init})$,  see Eq.~(\ref{critical})), since we are interested in the early 2-2-domination scenario. From the condition $M_{\bar{\textrm{c}}}\lesssim \Mini\lesssim M_\textrm{BBN}$, we can also determine the interval for the background temperature at the time of 2-2-hole formation as
\begin{eqnarray}
\label{Tinitalinterval}
3.6\times 10^{11} \hMmin^{-1/3}\;\rm{GeV}\lesssim T_{\mathrm{init}} \lesssim 2.4\times 10^{12} \hMmin^{-2/5}\;\rm{GeV}\;.
\end{eqnarray}
From Eqs.~(\ref{reheatinginterval}) and (\ref{Tinitalinterval}), we can also determine the upper limit for the remnant mass $\Mmin$, but the more stringent bound comes from the fact that $M_{\bar{\textrm{c}}}$ must not exceed $M_\textrm{BBN}$. Therefore, in the early 2-2-hole domination scenario, we have the following condition for $\Mmin$.
\begin{eqnarray}
\label{Mmininterval}
\Mp\; \lesssim \Mmin \lesssim 3.3\times 10^{7}\; \rm{g}\;.
\end{eqnarray}
However, this can be improved further. Ref.~\cite{Domenech:2020ssp} investigated the BBN constraints on the energy density of the gravitational waves produced at the time when primordial black holes, dominating the energy density, evaporated and reheated the universe into a radiation-dominated phase. This leads to a constraint on the initial fraction of the holes~\cite{Domenech:2020ssp},
\begin{eqnarray}
\label{betamax}
\beta < 1.1\times 10^{-6} \left(\frac{\gamma}{0.2}\right)^{-1/2}\left(\frac{\mathcal{N}_*}{108}\right)^{17/48}\left(\frac{g_{*}(T_{\tau})}{106.75}\right)^{1/16}\left(\frac{\Mini}{10^{4}\rm{g}}\right)^{-17/24}\equiv \beta_{\rm{max}}\;.
\end{eqnarray}
This should also be valid for 2-2-holes. The formation and almost instantaneous evaporation of 2-2-holes, in theory, is almost identical to the BH case (as discussed in Appendix~\ref{22holereview}). The difference is that 2-2-holes leave behind remnants, which seemingly do not play a role in the calculation of the gravitational wave production after evaporation. Since these are the main ingredients of the analysis of Ref.~\cite{Domenech:2020ssp}\footnote{Note that  Ref.~\cite{Domenech:2020ssp}'s analysis is for PBHs with monochromatic mass function at formation, which is known to be a well-working approximation to constraint the parameter space~\cite{Carr:2009jm} and which we adopt in this paper for 2-2-holes as well. For a recent discussion of PBHs with extended mass functions, see Ref.~\cite{Papanikolaou:2022chm}.}, we simply assume that the bound on the initial fraction, given in Eq.~(\ref{betamax}), applies to 2-2-holes as well. Note that 

Then, with our assumptions of $\gamma\simeq 0.2$, $g_{*}(T_{\tau})\simeq 12$, and $\mathcal{N}_* \simeq 108$,  the 2-2-domination condition ($\beta\gtrsim \beta_c$, where $\beta_c$ is given in Eq.~(\ref{critical})), together with Eq.~(\ref{betamax}), leads to the condition that $\Mmin \lesssim 43\;\rm{g}\;$. Furthermore, it is of course required that the background temperature at around which 2-2-holes start to dominate the energy budget, which will be denoted as $T_{\rm{dom}}$ (see Eq.~(\ref{Tdom})), is larger than the evaporation temperature of 2-2-holes $T_{\tau}$ (given in Eq.~(\ref{reheatingtemp}). The necessary (but not sufficient condition) for $T_{\rm{dom}}\gtrsim T_{\tau}$ becomes $\Mmin \lesssim 0.1\;\rm{g}\;$, which is obtained by choosing  $\beta=\beta_{\rm{max}}$ and  $\Mini=M_{\rm{BBN}}$. Therefore, our allowed interval for the remnant mass is finally given as 
\begin{eqnarray}
\label{Mminintervalfinal}
\Mp=2.2\times 10^{-5}\;\rm{g}\; \lesssim \Mmin \lesssim 0.1\; \rm{g}\;.
\end{eqnarray}
%


\section{Effects of 2-2-hole entropy injection \label{sec:entropy-injection}}
Here, we look at the effects of entropy injection due to the 2-2-hole evaporation on the evolution of the universe and axion abundance. This is in analogy to the PBH case discussed in Ref.~\cite{Bernal:2021yyb}.
Considering that we have now a 2-2-hole dominant stage (i.e. $n(t_{\mathrm{init}})\gtrsim n_c$), at some point in time, say $t=t_{\mathrm{eq}}$, the radiation and 2-2-hole densities become equal; at a later time, say $t_\mathrm{dom}$, the 2-2-hole domination begins and lasts till the time 2-2-holes become remnants, i.e. at $t\approx t_{\mathrm{init}}+\tau_L\approx \tau_L$. The formation time ($t_{\mathrm{init}}$) and evaporation time ($t_{\mathrm{eva}}\approx \tau_L$) (see Eq.~(\ref{eq:tauL})) are given as 
\begin{eqnarray}  
\label{timevalues}  
\frac{t_{\mathrm{init}}}{\mathrm{s}}=10^{-38}  \;\frac{M_{\mathrm{init}}}{\mathrm{g}}\;,\qquad \qquad  \frac{\tau_L}{\mathrm{s}}=2\times 10^{-26}\hat{M}_{\mathrm{min}}^{-2}\left(\frac{M_{\mathrm{init}}}{\mathrm{g}}\right)^3\;.
\end{eqnarray}

 Note that we do not want to disrupt BBN. Therefore, whatever happens should happen before BBN, i.e. $t\lesssim1$ s or  $T(\tau_L) \gtrsim 1$ MeV.  This would require the aforementioned condition on the initial (or formation) mass of the 2-2-holes  $\Mini\lesssim M_\textrm{BBN}$, where the definition of the latter is given in Eq.~(\ref{MBBN}).
 
 Since the 2-2-hole domination in our scenario occurs before $t\sim 10^{-2}$ s, it lasts too short to have any noticeable effects regarding the timing and duration of the significant processes in the evolution of the universe. Hence, for instance, it cannot directly solve the Hubble tension problem. One possible way that could affect this is the production of extra radiation degrees of freedom, namely changing $N_{\mathrm{eff}}$.~In our scenario, since the only non-SM degree of freedom is the axion, there is no significant difference from the standard scenario~\cite{Bernal:2021yyb}, as opposed to the case where we may have a large dark sector~\cite{Aydemir:2020pao}.
 
On the other hand, such a change in the evolution of the universe can affect the parameter space of the axion DM scenario, which will be finalized later in this section.

 \subsection{Evolution of the very early universe}
Due to the intermediate era of 2-2-hole domination,  the entropy injection from the 2-2-hole radiation cannot be ignored and therefore the total entropy is not conserved.  As a result, the simple scale-factor dependence of energy densities cannot be applied. Instead, the Boltzmann equations for the evolution of each component should be solved. Namely, we have the equations~\cite{Hooper:2019gtx}   
\begin{eqnarray}
\label{Boltzmanneqs}
\begin{aligned}
\frac{d \rho}{d t}+3 H \rho & =+\frac{\rho }{M} \frac{d M}{d t} \\
\frac{d \rho_R}{d t}+4 H \rho_R & =-\frac{\rho }{M} \frac{d M}{d t}\;,
\end{aligned}
\end{eqnarray}
 where $\rho$ and $\rho_R$ are the energy density of 2-2-holes and radiation. From the Friedmann equation, we have $H^2=8\pi (\rho_R+\rho)/(3\Mp^2)$. The axion is always subdominant for the time scale of interest. The radiation rate of axions from 2-2-holes is negligible. Therefore, the axion evolution equation is decoupled from the equations above and can be solved separately~\cite{Bernal:2021yyb,Arias:2021rer}. In fact, its evolution equation can be approximated as $d\rho_a/dt+ 3H\rho_a\simeq 0$ and hence it will always act like NR matter, i.e. $\rho_a\propto a^{-3}$. However, due to the time-dependent source terms on the RHS of Eq.~(\ref{Boltzmanneqs}), $\rho_R$ and $\rho$ do not always evolve as free-falling fluids. In order to find the full evaluation of these components, one has to solve these equations numerically, but approximate analytic solutions also are known to give good results~\cite{Bernal:2021yyb,Arias:2021rer,Arias:2020qty}, which we will be using below. Meanwhile, we will always have $\rho_R(T)=\frac{\pi^2}{30}g_*(T) T^4$  and will expressed our results in terms of $H_R\equiv \sqrt{8\pi \rho_R/(3\Mp^2)}$.
 
  As mentioned above, we have four regimes before BBN denoted as $t\leqslant t_{\mathrm{eq}}$, $t_{\mathrm{eq}}< t \leqslant t_{\mathrm{dom}}$, $t_{\mathrm{dom}}< t \leqslant \tau_L$, and $\tau_L <t$. Similar to the BH case~\cite{Bernal:2021yyb}, we can find the Hubble rate for the 2-2-hole case in each regime in terms of temperature as
  \begin{equation}
  \label{Hubblerates}
H(T) \simeq 
\begin{cases}H_R(T) & \text { for } T \geq T_{\mathrm{eq}}, 
\vspace{0.2cm}\\ H_R\left(T_{\mathrm{eq}}\right)\left[\dfrac{g_{* s}(T)}{g_{* s}\left(T_{\mathrm{eq}}\right)}\left(\dfrac{T}{T_{\mathrm{eq}}}\right)^3\right]^{1 / 2} & \text { for } T_{\mathrm{eq}} \geq T \geq T_{\mathrm{dom}}\;,
 \vspace{0.3cm}\\  H_R\left(T_{\tau}\right)\left[1-\dfrac{720\;(8\pi)^2}{\pi \;\mathcal{N}_*} \dfrac{\hat{M}_{\mathrm{init}}^3}{\mathcal{A}^4} \dfrac{H_R^2\left(T_{\tau}\right)-H_R^2(T)}{\Mp\;H_R\left(T_{\tau}\right)}\right] & \text { for } T_{\mathrm{dom}} \geq T \geq T_{\tau}\;,
 \vspace{0.2cm} \\ H_R(T) & \text { for } T_{\tau} \geq T\;,
 \end{cases}
\end{equation}  
where $\mathcal{A}=1.7\, \mathcal{N}^{-1/4}\hMmin^{1/2}$, introduced in Eq.~(\ref{22holetemp}).
In contrast to the BH case, in the third regime, where 2-2-hole domination occurs, we have a dependence on our extra parameter $\hMmin\equiv\Mmin/\Mp$, the remnant mass, in addition to the initial 2-2-hole mass, $\hat{M}_{\mathrm{init}}\equiv\Mini/\Mp$ through the dependence on $\mathcal{A}$. Taking $\mathcal{A}\rightarrow 1$ brings us to the BH case. We also recall that this is different from a possible BH remnant case as well since the BH remnants are generally anticipated to have Planck mass, whereas $\Mmin$ does not have a theoretical upper bound. Notice that the reheating temperature $T_{\tau}$, which appears in the third interval, includes factors of $\Mmin$ and $\Mini$, as can be seen in Eq.~(\ref{reheatingtemp}).  In this interval, the denominator of the second term, in addition to $\mathcal{N}_*$, has also a factor of  $\mathcal{N}^{-1}$, contained in the $\mathcal{A}^4$ factor.  Recall that for our case, $\mathcal{N}=\mathcal{N}_*\approx 108$, and thus, their effects cancel in the quantities in question. What is important here is that the $\mathcal{N}_*$ factor also exists in the BH case since it represents the degrees of freedom (dofs) radiated at the initial temperature of the hole, and for such small holes $\mathcal{N}_*$ takes the maximum value, which is the value above if we assume only the SM dofs. But here we also have the factor $\mathcal{N}$, corresponding to the dofs within the 2-2-hole, which also gets extremely (actually divergently) hot deep inside. In most of the cases~\cite{Aydemir:2020xfd,Aydemir:2020pao}, most of the quantities that matter come with small exponents such as $\pm 1/4$ and thus do not play a big role. But if they come with higher exponents as in above this may lead to noticeable effects. Overall, it could be better to approximate the Hubble rates, given in Eq.~(\ref{Hubblerates}), further in order to compare and contrast with the black hole case (through the $\Mmin$ dependence) as 
  \begin{equation}
  \label{Hubblerates2}
\dfrac{H(T)}{\rm{GeV}} \simeq 
\begin{cases}1.4\times 10^{-18}\left(\dfrac{T}{\rm{GeV}}\right)^2 & \text { for } T \geq T_{\mathrm{eq}}, 
\vspace{0.2cm}\\ 9.3\times 10^{-11} \;\beta^{1/2} \left(\dfrac{\Mini}{\rm{gr}}\right)^{-1/4}\left(\dfrac{T}{\rm{GeV}}\right)^{3/2}& \text { for } T_{\mathrm{eq}} \geq T \geqslant T_{\mathrm{dom}}\;,
 \vspace{0.3cm}\\  2.6\times 10^{-37}\;\hMmin^{-2} \left(\dfrac{\Mini}{\rm{gr}}\right)^{3}\left(\dfrac{T}{\rm{GeV}}\right)^{4} & \text { for } T_{\mathrm{dom}} \geq T \gg T_{\tau}\;,
 \vspace{0.2cm} \\ 4.7\times 10^{-19}\left(\dfrac{T}{\rm{GeV}}\right)^2 & \text { for } T_{\tau} \geq T\;.
 \end{cases}
\end{equation}  
Notice that in the second expression, the initial fraction of 2-2-holes, $\beta$, appears through the definition of $T_{\mathrm{eq}}$, given in Eq.~(\ref{Teq}). 

By matching the second and third lines in Eq.~(\ref{Hubblerates}) at $T=T_{\rm{dom}}$, we can estimate the background temperature at around which 2-2-holes start to dominate the energy budget as
\begin{eqnarray}
\label{Tdom}
T_{\rm{dom}}&\simeq& \left[ \left(\frac{\Mp}{\sqrt{8\pi}}\right)^{10} \frac{\mathcal{A}^8}{\Mini^6}\frac{\mathcal{N}_*^2}{5760\;g_*(T_{\rm{dom}})}T_{\mathrm{eq}}\right]^{1/5}\nonumber\\
&=& 2.3\times 10^{7}\;\hMmin^{4/5}\;\left(\frac{M_{\mathrm{init}}}{\mathrm{g}}\right)^{-6/5} \;\left(\frac{T_{\rm{eq}}}{\rm{GeV}}\right)^{1/5}\rm{GeV}\nonumber\\
&=& 3.6\times 10^{10}\;\beta^{1/5}\;\hMmin^{4/5} \left(\frac{M_{\mathrm{init}}}{\mathrm{g}}\right)^{-13/10}\rm{GeV}\;,
\end{eqnarray}
 where we use Eq.~(\ref{Teq}) and take into account that $H_R(T_{\rm{dom}})\gg H_R(T_{\tau})$ (since $T_{\rm{dom}} \gg T_{\tau}$). 
 
The entropy injection factor can also be found, similar to the BH case~\cite{Bernal:2021yyb}, as
 \begin{equation}
  \label{entropyinjection} 
\frac{S(T)}{S(T_{\tau})} \simeq 
\begin{cases}\dfrac{g_{*,s}(T_{\rm{eq}})}{g_{*,s}(T_{\tau})}\dfrac{g_{*}(T_{\tau})}{g_{*}(T_{\rm{eq}})}\dfrac{T_{\tau}}{T_{\rm{eq}}} & \text { for }T \geq T_{\mathrm{dom}}\;,
 \vspace{0.3cm}\\  \dfrac{g_{*,s}(T)}{g_{*,s}(T_{\tau})}\left(\dfrac{T}{T_{\tau}}\right)^3 \left[1-\dfrac{720\;(8\pi)^2}{\pi \;\mathcal{N}_*} \dfrac{\hat{M}_{\mathrm{init}}^3}{\mathcal{A}^4} \dfrac{H_R^2\left(T_{\tau}\right)-H_R^2(T)}{\Mp\;H_R\left(T_{\tau}\right)}\right]^{-2} & \text { for } T_{\mathrm{dom}} \geq T \geq T_{\tau}\;,
 \vspace{0.2cm} \\ 1 & \text { for } T_{\tau} \geq T\;,
 \end{cases}
\end{equation}  
where there are three regimes, instead of the four in Eq.~(\ref{Hubblerates}), due to the approximate entropy conservation for $ T\geqslant T_{\rm{eq}}$. Recall that $g_{*,s}\approx g_*$ at temperature values of interest, i.e. above BBN scale~\cite{Husdal:2016haj}. As in the case of Hubble rates, we can simplify these expressions further as
 \begin{equation}
  \label{entropyinjection2} 
\frac{S(T)}{S(T_{\tau})} \simeq 
\begin{cases}1.6\times 10^{-6}\;\beta^{-1} \hMmin\left(\dfrac{M_{\mathrm{init}}}{\mathrm{g}}\right)^{-1} & \text { for }T \geq T_{\mathrm{dom}}\;,
 \vspace{0.3cm}\\  1.8\times 10^{47}\;\hMmin^5\left(\dfrac{M_{\mathrm{init}}}{\mathrm{g}}\right)^{-15/2}\left(\dfrac{T}{\rm{GeV}}\right)^{-5} & \text { for } T_{\mathrm{dom}} \geq T \gg T_{\tau}\;,
 \vspace{0.2cm} \\ 1 & \text { for } T_{\tau} \geq T\;.
 \end{cases}
\end{equation}  

 \subsection{Relation to the axion abundance}
 The axion abundance is given as 
 \begin{eqnarray}
 \label{axionabundance1}
\Omega_a h^2=\frac{\rho_a(T_0)}{\rho_c/h^2}=\frac{\rho_a(T_{\mathrm{osc}})}{\rho_c/h^2} \;\frac{m_a}{\tilde{m}_a(T_{\mathrm{osc}})} \frac{s(T_0)}{s(T_{\mathrm{osc}})}\frac{S(T_{\mathrm{osc}})}{S(T_\tau)}\;,
\end{eqnarray}
where $\rho_a(T_{\mathrm{osc}})\approx\frac{1}{2} \tilde{m}_a^2(T_{\mathrm{osc}}) f_a^2\; \theta_i^2$ is the axion energy density at the oscillation temperature and $s(T)=\frac{2\pi^2}{45}g_* T^3$ is the total entropy density. The entropy density and the critical energy density at present are given as $s(T_0)=2.7\times 10^3\;\mathrm{cm}^{-3}$ and $\rho_c/h^2=1.1\times 10^{-5} \;\mathrm{GeV}\;\mathrm{cm}^{-3}$, respectively.  Therefore, the abundance becomes
\begin{eqnarray}
\label{axionabundance2}
\Omega_a h^2\simeq 2.9\times 10^3 \;\theta_i^2\; g_{*,s}^{-1}\left(T_{\mathrm{osc}}\right) \; \frac{\tilde{m}_a(T_{\mathrm{osc}})}{m_a}\left(\frac{T_{\mathrm{osc}}}{\mathrm{GeV}}\right)^{-3}\times \frac{S(T_{\mathrm{osc}})}{S(T_\tau)}\;.
\end{eqnarray}
Since $T_{\mathrm{osc}}$ is determined by $3 H (T_{\mathrm{osc}})= \tilde{m}_a (T_{\mathrm{osc}})$, from Eqs.~(\ref{eqn:m(T)}) and (\ref{Hubblerates2}), we find for $T_{\rm{osc}}\leqslant T_{\rm{QCD}}$
 \begin{equation}
  \label{Tosc1}
\dfrac{T_{\rm{osc}}}{\rm{GeV}} \simeq 
\begin{cases}1.5\times 10^{4}\left(\dfrac{m_a}{\rm{eV}}\right)^{1/2} & \text { for } T _{\rm{osc}} \geq T_{\mathrm{eq}}, 
\vspace{0.2cm}\\ 2.3 \;\beta^{-1/3} \left(\dfrac{\Mini}{\rm{gr}}\right)^{1/6}\left(\dfrac{m_a}{\rm{eV}}\right)^{2/3}& \text { for } T_{\mathrm{eq}} \geq T_{\rm{osc}} \geqslant T_{\mathrm{dom}}\;,
 \vspace{0.3cm}\\  6.0\times 10^{6}\;\hMmin^{1/2} \left(\dfrac{\Mini}{\rm{gr}}\right)^{-3/4}\left(\dfrac{m_a}{\rm{eV}}\right)^{1/4} & \text { for } T_{\mathrm{dom}} \geq T_{\rm{osc}} \gg T_{\tau}\;,
 \vspace{0.2cm} \\ 2.7\times 10^{4}\left(\dfrac{m_a}{\rm{eV}}\right)^{1/2} & \text { for } T_{\tau} \geq T_{\rm{osc}}\;,
 \end{cases}
\end{equation}  
and for $T_{\rm{osc}}\gtrsim T_{\rm{QCD}}$
 \begin{equation}
  \label{Tosc2}
\dfrac{T_{\rm{osc}}}{\rm{GeV}} \simeq 
\begin{cases}7.0\left(\dfrac{m_a}{\rm{eV}}\right)^{1/6} & \text { for } T _{\rm{osc}} \geq T_{\mathrm{eq}}, 
\vspace{0.2cm}\\ 0.34 \;\beta^{-1/11} \left(\dfrac{\Mini}{\rm{gr}}\right)^{1/22}\left(\dfrac{m_a}{\rm{eV}}\right)^{2/11}& \text { for } T_{\mathrm{eq}} \geq T_{\rm{osc}} \geqslant T_{\mathrm{dom}}\;,
 \vspace{0.3cm}\\  9.5\times 10^{2}\;\hMmin^{1/4} \left(\dfrac{\Mini}{\rm{gr}}\right)^{-3/8}\left(\dfrac{m_a}{\rm{eV}}\right)^{1/8} & \text { for } T_{\mathrm{dom}} \geq T_{\rm{osc}} \gg T_{\tau}\;,
 \vspace{0.2cm} \\ 8.4\left(\dfrac{m_a}{\rm{eV}}\right)^{1/6} & \text { for } T_{\tau} \geq T_{\rm{osc}}\;,
 \end{cases}
\end{equation}  
where we recall that $T_{\rm{QCD}}\simeq150$ MeV. Then, from Eqs.~(\ref{axionabundance2}), (\ref{Tosc1}), (\ref{entropyinjection2}), and (\ref{eqn:m(T)}), we find the axion abundance  today, for $T_{\rm{osc}}\leqslant T_{\rm{QCD}}$, as
\begin{equation}
  \label{axionabundance3}
\dfrac{\Omega_a h^2}{0.12} \simeq 
\begin{cases}\;\theta_i^2\;\left(\dfrac{\beta}{10^{-13}}\right)^{-1}\left(\dfrac{m_a}{0.4\times 10^{-7}\;\rm{eV}}\right)^{-\frac{3}{2}}\hMmin \left(\dfrac{\Mini}{10^{8}\;\rm{g}}\right)^{-1} & \text { for } T _{\rm{osc}} \geq T_{\mathrm{eq}}, 
\vspace{0.2cm}\\ \;\theta_i^2 \;\left(\dfrac{m_a}{0.5\times 10^{-8}\;\rm{eV}}\right)^{-2}\;\hMmin\;\left(\dfrac{\Mini}{10^8\;\rm{g}}\right)^{-\frac{3}{2}}& \text { for } T_{\mathrm{eq}} \geq T_{\rm{osc}} \geqslant T_{\mathrm{dom}}\;,
 \vspace{0.3cm}\\ \; \theta_i^2 \;\left(\dfrac{m_a}{0.5\times 10^{-8}\;\rm{eV}}\right)^{-2}\;\hMmin\;\left(\dfrac{\Mini}{10^8\;\rm{g}}\right)^{-\frac{3}{2}}& \text { for } T_{\mathrm{dom}} \geq T_{\rm{osc}} \gg T_{\tau}\;,
 \vspace{0.2cm} \\ \;\theta_i^2\;\left(\dfrac{m_a}{4.4\times10^{-7}\;\rm{eV}}\right)^{-\frac{3}{2}}& \text { for } T_{\tau} \geq T_{\rm{osc}}\;,
 \end{cases}
\end{equation}  
and for $T_{\rm{osc}}\gtrsim T_{\rm{QCD}}$ as

\begin{equation}
  \label{axionabundance4}
\dfrac{\Omega_a h^2}{0.12} \simeq 
\begin{cases}\;\theta_i^2\;\left(\dfrac{\beta}{10^{-13}}\right)^{-1}\left(\dfrac{m_a}{3.7\times 10^{-7}\;\rm{eV}}\right)^{-\frac{7}{6}}\hMmin \left(\dfrac{\Mini}{10^{8}\;\rm{g}}\right)^{-1} & \text { for } T _{\rm{osc}} \geq T_{\mathrm{eq}}, 
\vspace{0.2cm}\\ \;\theta_i^2\;\left(\dfrac{\beta}{10^{-13}}\right)^{-\frac{4}{11}} \;\left(\dfrac{m_a}{1.4\times10^{-7}\;\rm{eV}}\right)^{-\frac{14}{11}}\;\hMmin\;\left(\dfrac{\Mini}{10^8\;\rm{g}}\right)^{-\frac{29}{22}}& \text { for } T_{\mathrm{eq}} \geq T_{\rm{osc}} \geqslant T_{\mathrm{dom}}\;,
 \vspace{0.3cm}\\ \; \theta_i^2 \;\left(\dfrac{m_a}{2.0\times 10^{-9}\;\rm{eV}}\right)^{-\frac{3}{2}}\;\hMmin^2\;\left(\dfrac{\Mini}{10^8\;\rm{g}}\right)^{-3}& \text { for } T_{\mathrm{dom}} \geq T_{\rm{osc}} \gg T_{\tau}\;,
 \vspace{0.2cm} \\ \;\theta_i^2\;\left(\dfrac{m_a}{6.9\times10^{-6}\;\rm{eV}}\right)^{-\frac{7}{6}}& \text { for } T_{\tau} \geq T_{\rm{osc}}\;.
 \end{cases}
\end{equation}  
%

\section{Results and discussion\label{sec:results}}

\begin{figure}[h!]
\captionsetup[subfigure]{labelformat=empty}
\centering
\hspace{-0.8cm}
\begin{tabular}{lll}
\subfloat[(a)\quad  $\Mmin=3.6\;\Mp=7.9\times 10^{-5}\;\rm{g}$]{\label{axionmassa}\includegraphics[width=8.2cm]{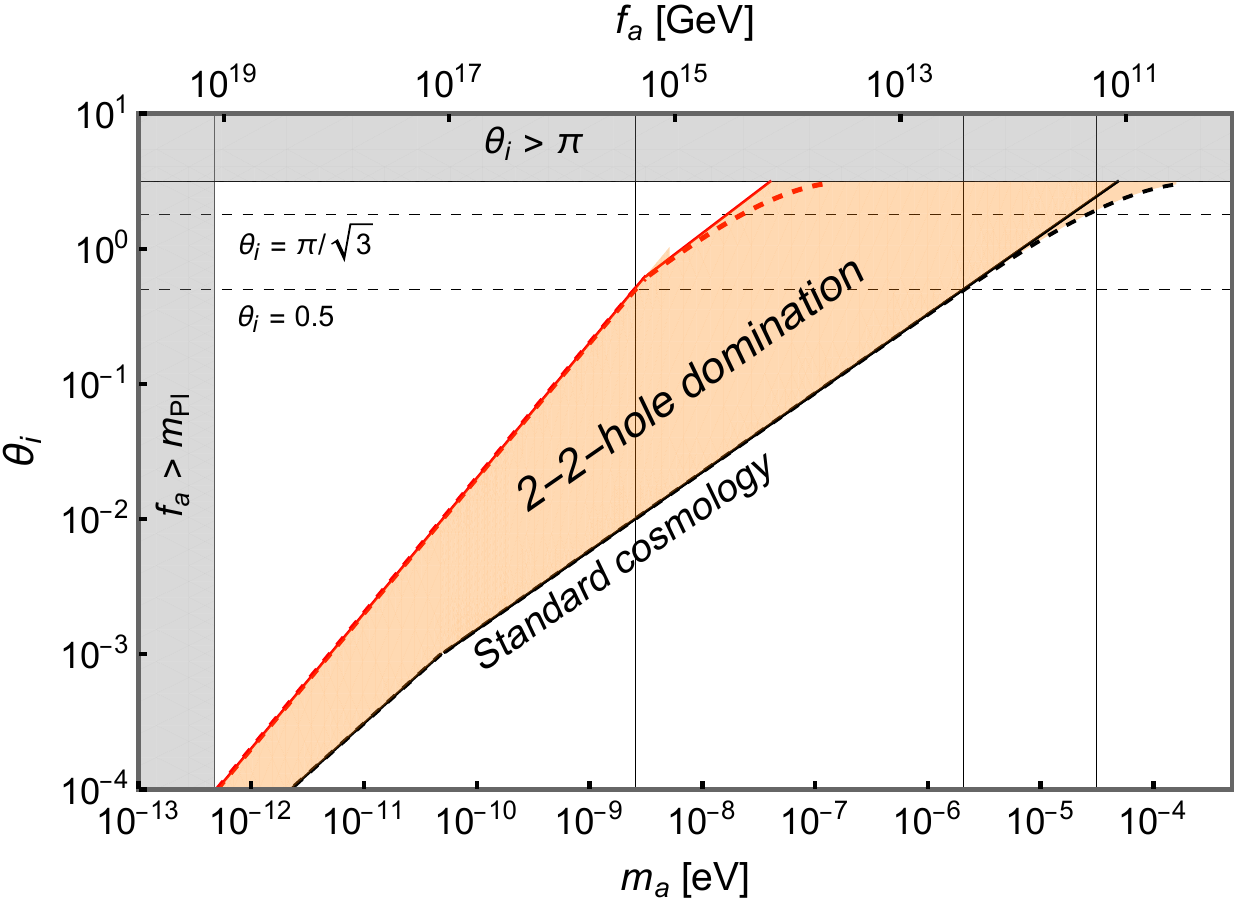}\label{plot1}}  \quad
\subfloat[(b)\quad   $\Mmin=0.1\;\rm{g}$]{\label{axionmassb}\includegraphics[width=8.2cm]{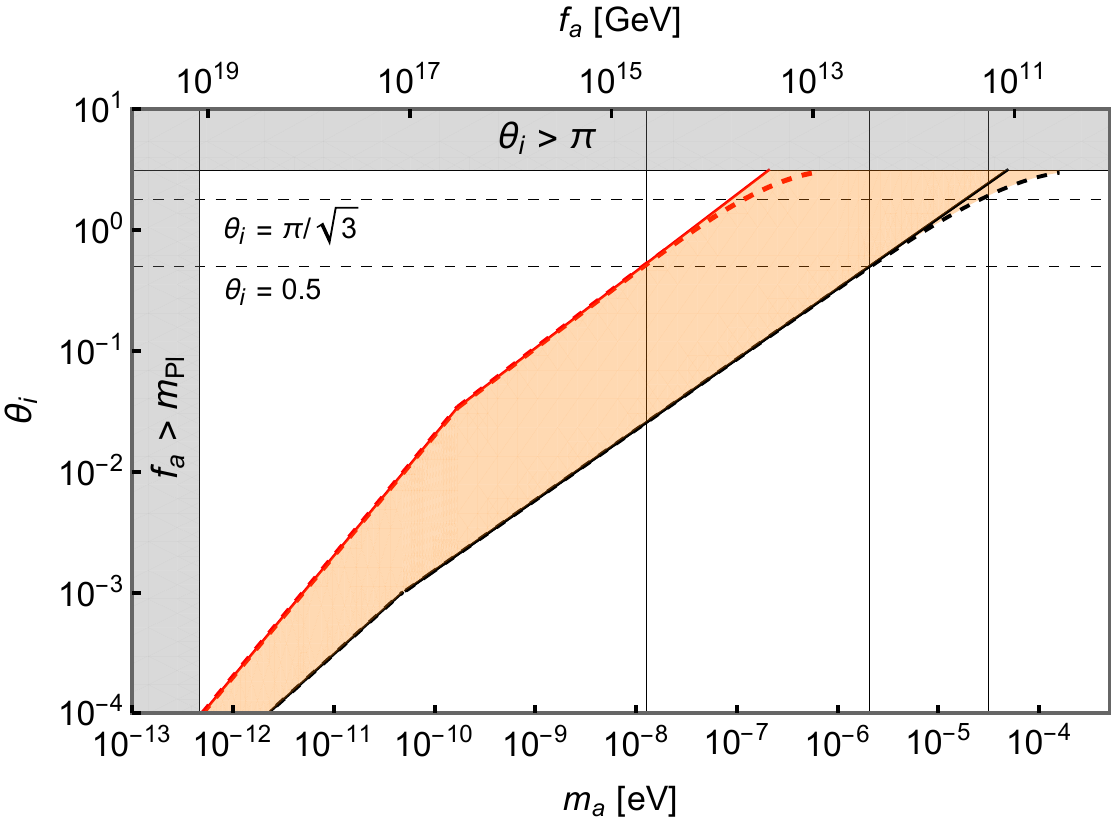}\label{plot2}}
\end{tabular}
\caption{\label{axionmass} 
The parameter space for the initial misalignment angle required for the QCD axion to generate all dark matter (modulo negligible contribution from 2-2-hole remnants) in the 2-2-hole domination scenario, given for two benchmark values of the remnant mass $\Mmin$,  corresponding to the (almost) minimum and maximum allowed values for the remnant mass in our scenario, as given in Eq.~(\ref{Mminintervalfinal}). The black lines correspond to the standard cosmological scenario, where there are no effects of 2-2-hole energy density in the evolution of the universe. The red lines denote the upper boundary ($\beta=\beta_{\rm{max}}$ and $\Mini=M_{\rm{BBN}}$) of the pre-BBN 2-2-hole domination scenario.  The gridlines display the intersection points with the desired $\theta_i$ values. The  $\Mmin=3.6\;\Mp$ case parametrically yields the black hole scenario and can be compared to the results of Ref.~\cite{Bernal:2021yyb}. Therefore, our results for $\Mmin=3.6\;\Mp$ can be considered the updated results for the black hole case since we use the most recent upper bound on $\beta$, given in Eq.~(\ref{betamax}). We also take into account the anharmonicity effects of the axion potential, denoted as dashed lines in the figure. The largest allowed $m_a$ interval for $0.5\leqslant \theta_i \leqslant \pi/\sqrt{3}$ is obtained in the $\Mmin=\Mp$ limit. The interval in Figure~\ref{plot1} is found as $2.5\times 10^{-9}\;\rm{eV}\leqslant m_a \leqslant 2.7\times 10^{-5}\;\rm{eV}$.}   
\end{figure}

Here, we first present the maximum parameter space for the misalignment angle of the QCD axion, required to generate the entire dark matter abundance (modulo the negligible 2-2-hole remnant contribution). Adopting a similar style to Ref.~\cite{Bernal:2021yyb}, we present the results in Figure~\ref{axionmass}. Differently, not only do we take into account the anharmonicity effects of the axion potential (dashed lines), but we also use the most recent expression for the upper bound on $\beta$ coming from the gravitational waves, given in a revised version of Ref.~\cite{Domenech:2020ssp} (see Eq.~(\ref{betamax})). As can be seen from Eqs.~(\ref{axionabundance3}) and (\ref{axionabundance4}), the maximum allowed $\theta_i$ is obtained for maximum values for $\beta$ and $\Mini$ for a given remnant mass $\Mmin$, i.e. $\beta=\beta_{\rm{max}}(\Mini)$ and $\Mini=M_{\rm{BBN}}$, given in Eqs.~(\ref{betamax}) and (\ref{MBBN}), respectively. This is displayed as red lines in Figure~\ref{axionmass}, whereas the standard cosmological scenario, where there is no 2-2-hole (etc.) domination era, is shown as black lines. The anharmonicity effects of the axion potential, as described in Ref.~\cite{Visinelli:2009zm}, are given as dashed lines. Notice that they are only effective for larger values of $\theta_i$, as expected.

Taking the optimal interval as $0.5\leqslant \theta_i \leqslant \pi/\sqrt{3}$ yields an allowed interval for the axion mass. As mentioned in Section \ref{sec:Axion-review}, this interval is not carved in stone; it is just taken as a favorable window, specifically for the angle not being "too small" since QCD axion attempts to explain the smallness of the $\theta$ parameter of QCD, namely the strong CP problem, in the first place. The largest range of allowed axion mass for this angle interval is obtained for $\Mmin=\Mp$ as 
\begin{eqnarray}
\label{mainterval}
1.9\times 10^{-9}\;\rm{eV}\lesssim m_a \lesssim2.7\times 10^{-5}\;\rm{eV.}
\end{eqnarray}
This is an improvement compared to the standard cosmological scenario, where the lower limit is  $2.1\times 10^{-6}\;\rm{eV}$, as can be seen from the black line in Figure~\ref{axionmassa}. Recall that our scenario in the $\Mmin\simeq\Mp$ limit parametrically reduces to the black hole remnant case with the Planck mass.\footnote{More precisely, in this limit, our case reduces to the black hole \textit{remnant} scenario with the Planck mass remnants. But, in "the axion as (almost) all dark matter scenario", as we will discuss below, we ensure that the remnant contribution to dark matter is negligible and hence remnants do not play a noticeable role.} This is realized when the factor $\mathcal{A}$, defined in Eq.~(\ref{22holetemp}), goes to unity, which happens when $\Mmin=3.6\;\Mp$ (for $\mathcal{N}=108$). Therefore, Figure~\ref{axionmassa}, displayed for this remnant mass value, can be considered the updated version for the black hole case given in Ref.~\cite{Bernal:2021yyb}, since we use the most recent gravitational wave bounds on the initial fraction of the holes, $\beta$, given in Ref.~\cite{Domenech:2020ssp}. We also take into account the anharmonicity effects through the anharmonicity factor, given in Eq.~(\ref{fann}).~These effects become only slightly noticeable for relatively larger values and they will not be considered for the rest of our analysis. The black lines denote the standard cosmological scenario, where there is no 2-2-hole/black hole domination ($\beta < \beta_c$). The red lines are the upper bound coming from the hole domination    
with maximum $\beta=\beta_{\rm{max}}$ value, given in Eq.~(\ref{betamax}). Since the overall dependence of $\theta$ on the $\Mini$ is directly proportional, we also take the largest value for the initial hole, for the pre-BBN scenario, for the upper bound, namely $\Mini=M_{\rm{BBN}}$, which is given in Eq.~(\ref{MBBN}). The most narrow range for the pre-BBN 2-2-hole domination comes from the largest allowed remnant mass, $\Mmin=0.1\;\rm{g}$, shown in Figure~\ref{axionmassb}.

\begin{figure}[h!]
\captionsetup[subfigure]{labelformat=empty}
\centering
\hspace{-0.8cm}
\begin{tabular}{lll}
\subfloat[(a)\label{fig:beta1}\quad  $m_a=5.7\times10^{-7}\;\rm{eV}$;\quad $\Mmin=3.6\;\Mp$  .]{\includegraphics[width=7.0cm]{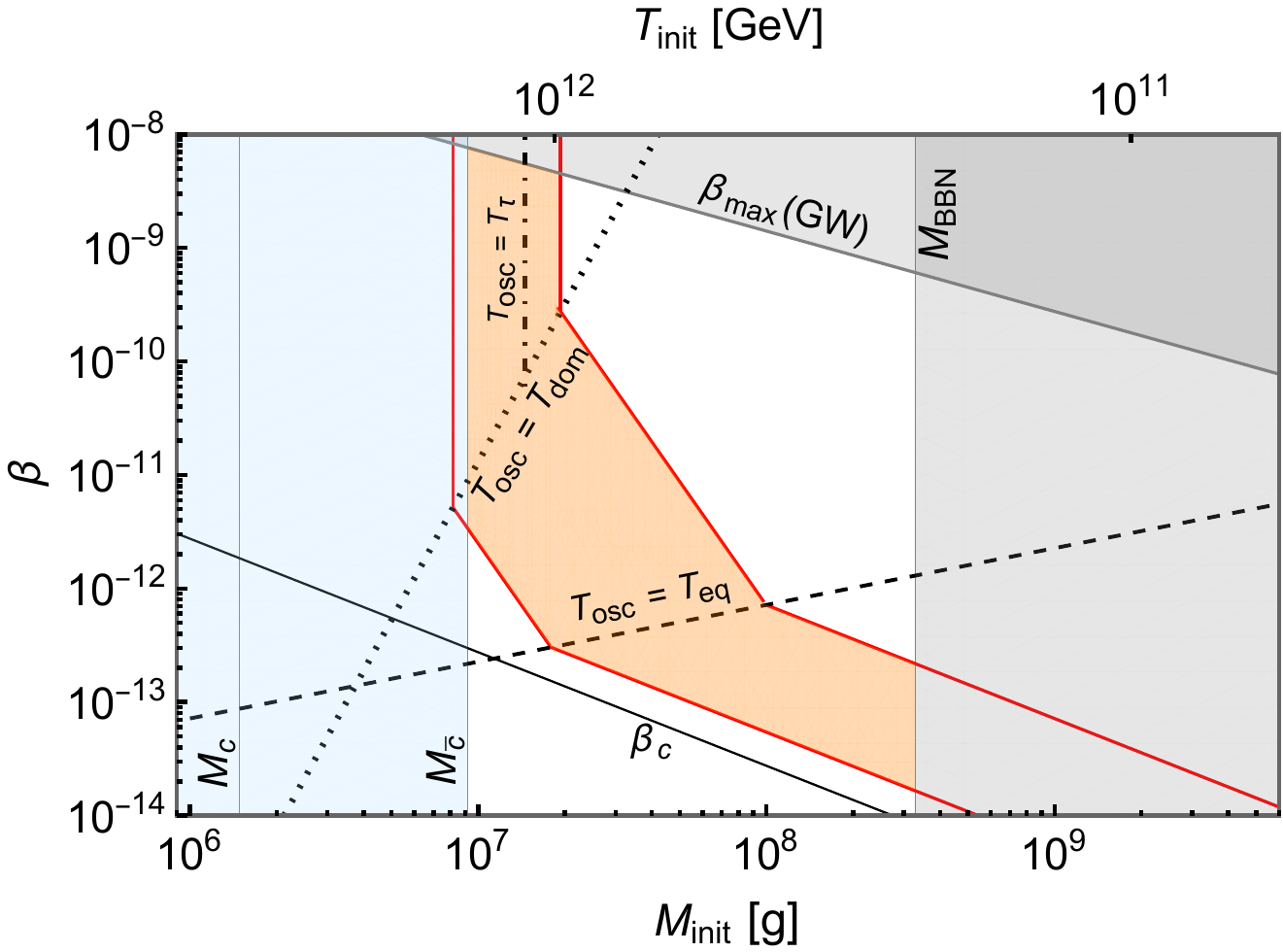}}  \quad
\subfloat[(b) \label{fig:beta2}\quad   $m_a=5.7\times10^{-7}\;\rm{eV}$;\quad $\Mmin=0.1\;\rm{g}$ .]{\includegraphics[width=6.9cm]{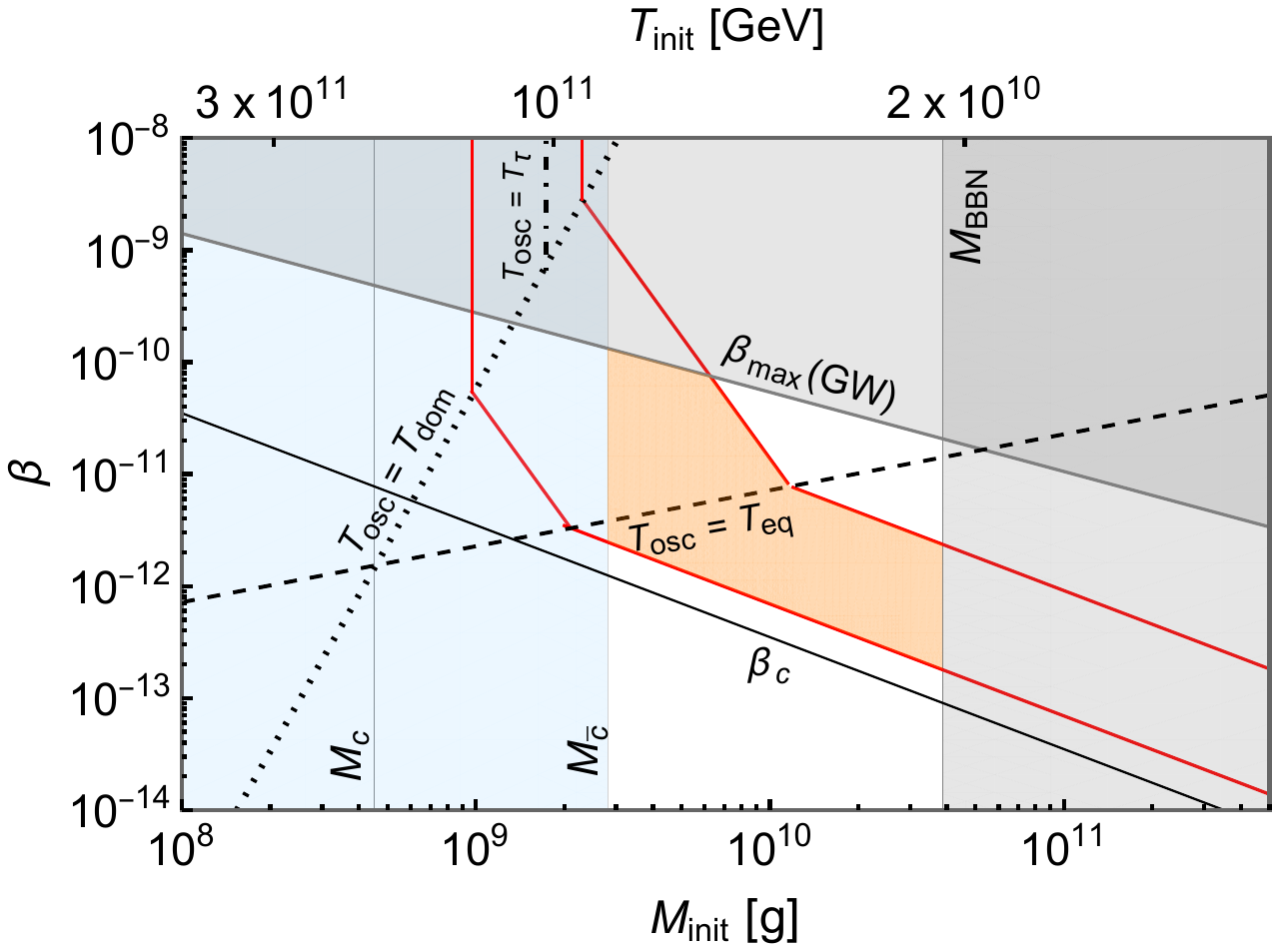}}\\\\
\subfloat[(c)\label{fig:beta3}\quad  $m_a=3.2\times10^{-9}\;\rm{eV}$;\quad $\Mmin=3.6\;\Mp$ .]{\includegraphics[width=6.95cm]{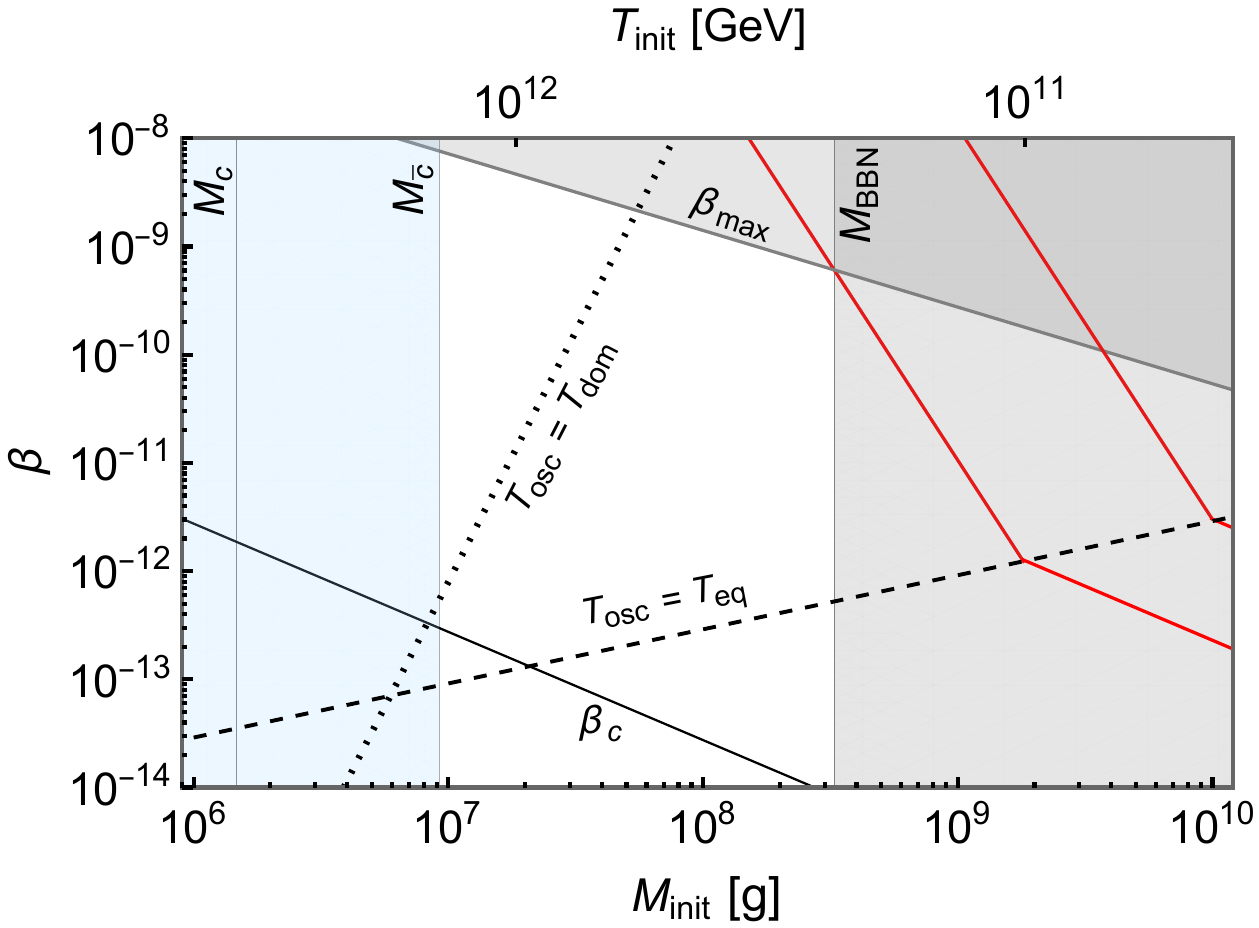}} \quad
\subfloat[(d)\label{fig:beta4}\quad  $m_a=5.7\times10^{-8}\;\rm{eV}$;\quad $\Mmin=0.1\;\rm{g}$ .]{\includegraphics[width=6.8cm]{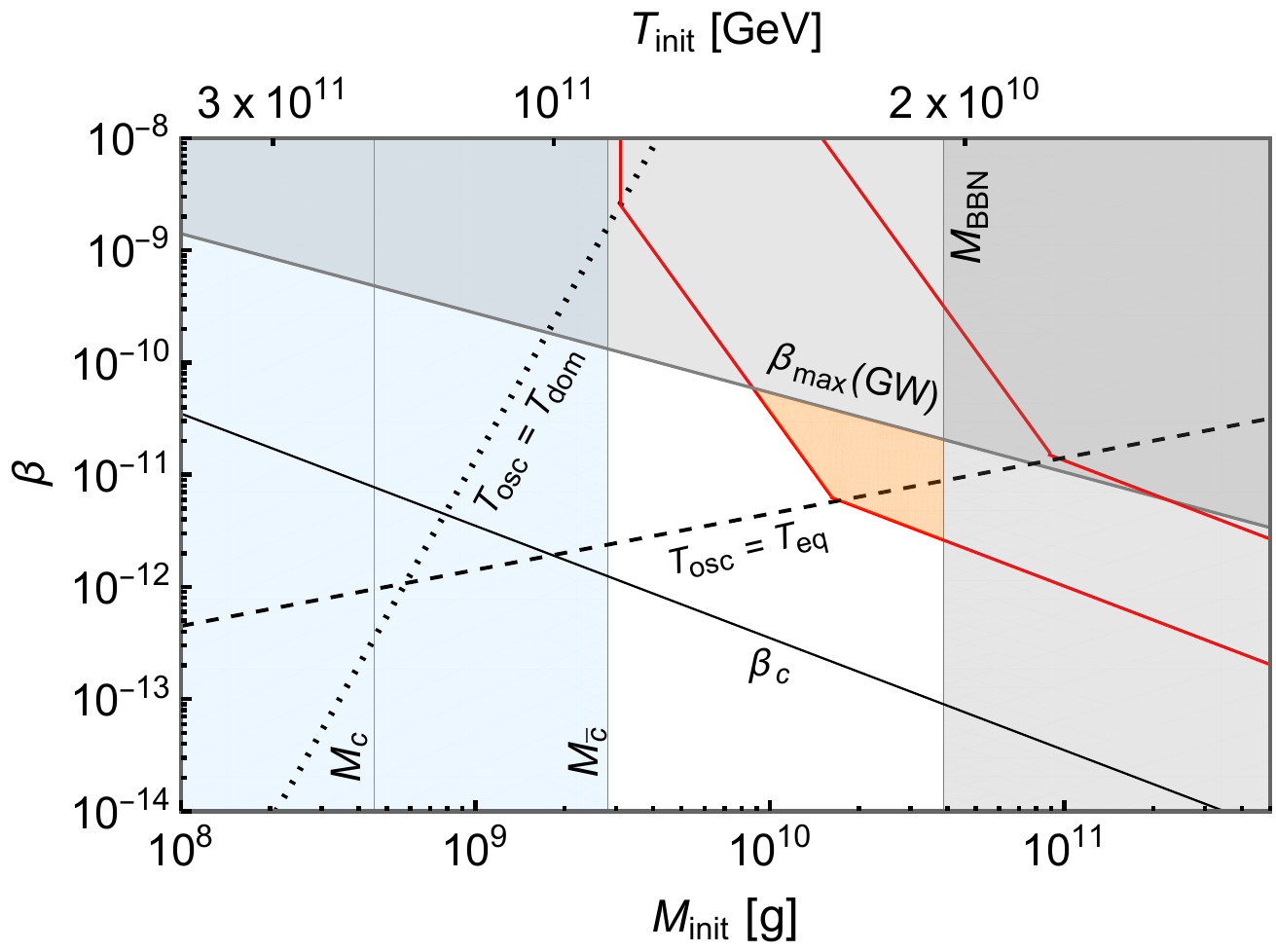}}
\end{tabular}
\caption{\label{fig:betalargeT} 
Constraints on the $\beta$ parameter, the initial fraction of 2-2-holes at formation, for $T_{\rm{osc}}\gtrsim T_{\rm{QCD}}$, for a given $\Mmin$. The region between red lines is the parameter space for the QCD axion to generate all dark matter (modulo negligible contribution from 2-2-hole remnants) in the 2-2-hole domination scenario for the interval  $0.5\leqslant \theta_i \leqslant \pi/\sqrt{3}$.  The small contribution to DM from remnants is chosen as one percent, which requires that $\Mini>M_{\bar{\textrm{c}}}$. We also display the critical mass $M_{\textrm{c}}$, indicating the case where the remnant would account for all DM. $\beta_c$ denotes the minimum possible value of $\beta$ required for the 2-2-hole domination to occur.  $\beta_{\rm{max}}$ is the upper bound coming from the gravitational wave constraints, investigated in Ref.~\cite{Domenech:2020ssp}. $M_{\rm{BBN}}$ is the upper value for $\Mini$ so that 2-2-holes become remnants before BBN begins. In the left column, we display the parameter space for $\Mmin=3.6\;\Mp$, which is equivalent to BHs with Planck mass remnants. For the regular BH scenario, where there are no BH remnants, the blue regions disappear and the parameter space in Figure~\ref{fig:beta1} gets slightly larger since small overlapping regions between the blue region and the red lines become also available. Note that the upper label in the frames denotes the background temperature in the universe at the time of hole formation.}   
\end{figure}

\begin{figure}[h!]
\captionsetup[subfigure]{labelformat=empty}
\centering
\hspace{-0.8cm}
\begin{tabular}{lll}
\subfloat[(a)\label{fig:beta5}\quad  $m_a=3.2\times10^{-9}\;\rm{eV}$;\quad $\Mmin=3.6\;\Mp$.]{\includegraphics[width=7.0cm]{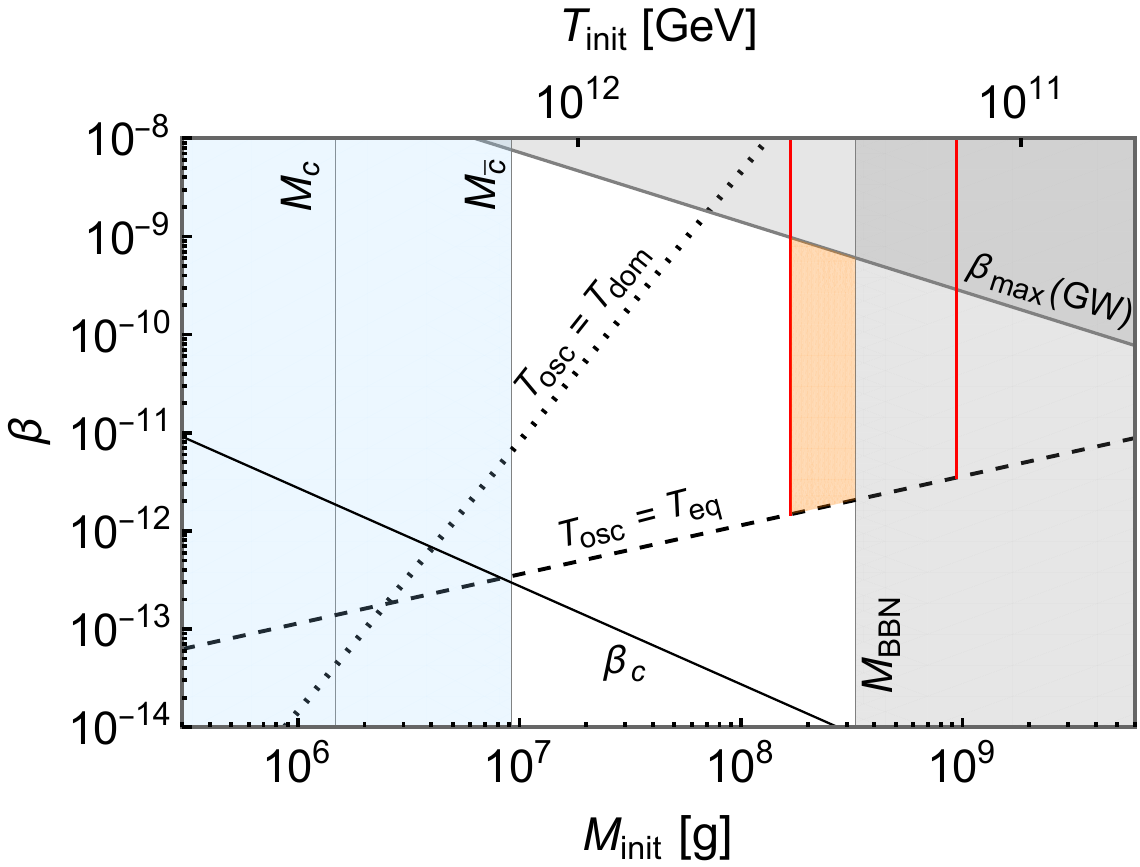}}  \quad
\subfloat[(b) \label{fig:beta6}\quad   $m_a=1.9\times10^{-9}\;\rm{eV}$;\quad $\Mmin=\Mp$ .]{\includegraphics[width=7.0cm]{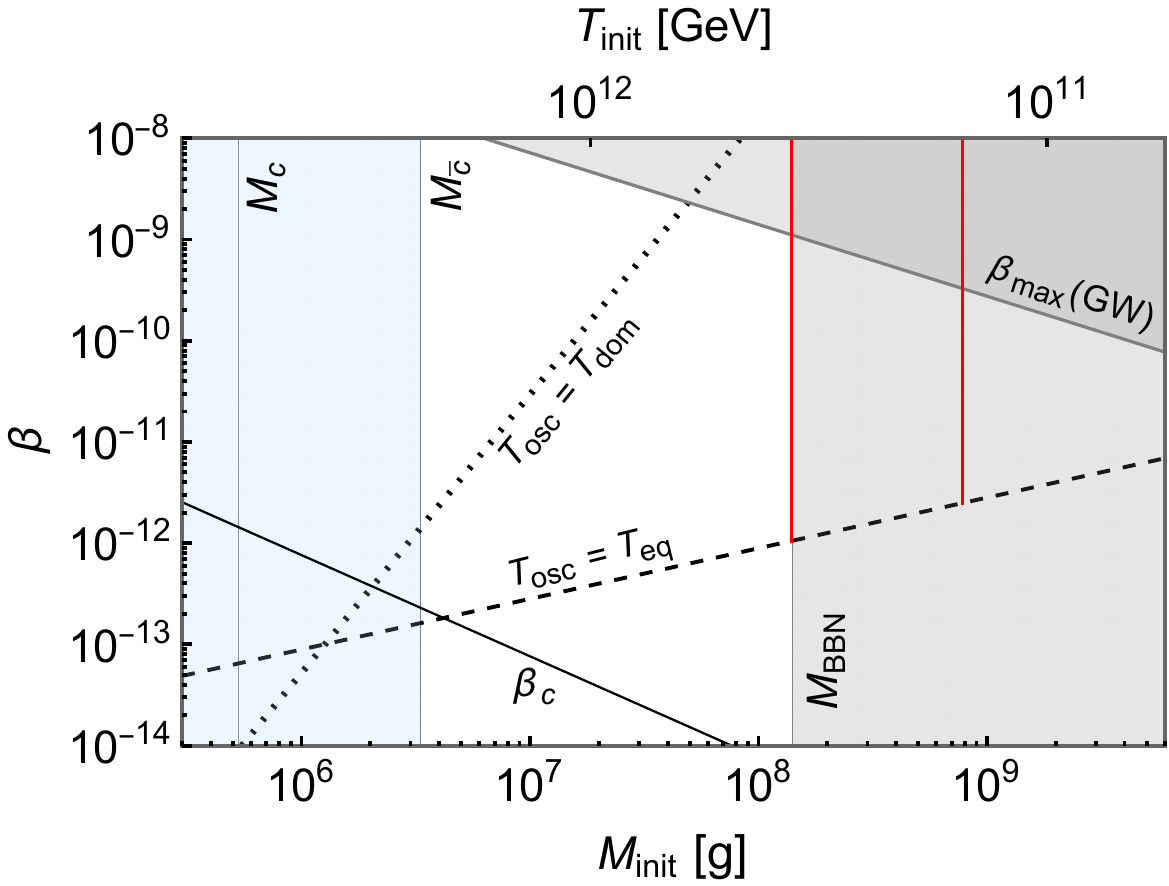}}
\end{tabular}
\caption{\label{fig:betasmallT} 
Constraints on the $\beta$ parameter for $T_{\rm{osc}}\lesssim T_{\rm{QCD}}$ for a given $\Mmin$. The only constraint on $\beta$, in this case, is that it is limited by the $\beta_{\mathrm{max}}$ and $T_{\rm{osc}}= T_{\rm{eq}}$ lines. As detailed in the text, the only relevant temperature interval here is $T_{\mathrm{eq}} \geqslant T_{\rm{osc}} \geqslant T_{\mathrm{dom}}$, where there is no $\beta$ dependence in the corresponding equation (given in Eq.~(\ref{axionabundance3})) and hence no constraint on $\beta$ in this region (as denoted by the vertical lines). In Figure~\ref{fig:beta5}, we again choose $\Mmin=3.6\;\Mp$, which reduces to the black hole case. 
 The knee point on Figure~\ref{plot1} is the intersection point between $\beta_{\mathrm{max}}$ and $M_{\rm{BBN}}$ lines on Figure~\ref{fig:beta5}. The smaller the axion mass, the smaller the allowed (orange) region. We display the smallest axion mass allowed (by our choice of angle interval $0.5\leqslant \theta_i \leqslant \pi/\sqrt{3}$) in Figure~\ref{fig:beta6}, which occurs fort the smallest 2-2-hole remnant mass, $\Mmin=\Mp$.}   
\end{figure}

In Figures~\ref{fig:betalargeT} and \ref{fig:betasmallT}, we display how the axion as all dark matter scenario (modulo negligible contribution from the remnants), can constrain the initial hole fraction, $\beta$, for the desired interval $0.5\leqslant \theta_i \leqslant \pi/\sqrt{3}$. In Figure~\ref{fig:betalargeT}, we present the result for the $T_{\rm{osc}}\gtrsim T_{\rm{QCD}}$ case. As done above, we choose one of our benchmark points as $\Mmin=3.6\;\Mp$ in Figure~\ref{fig:beta1}, which is equivalent to the black hole case (modulo the small overlap between blue and orange regions; see the next paragraph.) Our results are in slight disagreement with the results of Ref.~\cite{Bernal:2021yyb}, in regions that are not of interest (nonorange regions), which might be because they seem to be doing full numerical analysis, whereas we take into account the analytic results, given in Eqs.~(\ref{Tosc1}--\ref{axionabundance4}), since they work well in the region of interest (orange regions in the plots).

We also display the critical temperatures, discussed in the previous sections, in Figures~\ref{fig:betalargeT} and \ref{fig:betasmallT}. The lower parts of the plots denote higher temperatures. Notice that above the $T_{\rm{osc}}=T_{\rm{dom}}$ line, $\beta$ parameter is not constrained (the vertical red lines) since the abundance for  $T_{\rm{osc}}<T_{\rm{dom}}$ does not depend on $\beta$. In this region, the right side of the $T_{\rm{osc}}=T_\tau$ (dot-dashed) line denotes the interval $T_\tau< T_{\rm{osc}}<T_{\rm{dom}}$, whereas the left side of the line indicates the $T_{\rm{osc}}<T_\tau$ region.

Since we look for the pre-BBN scenario, $\Mini$ is bounded from above by $M_{\rm{BBN}}$ (given in Eq.~(\ref{MBBN})), so that, by the time BBN begins, the 2-2-holes become remnants (or, in the case of black holes, they complete their evaporation and disappear) and hence do not interfere with the BBN physics. The critical initial 2-2-hole mass, $M_{\bar{\textrm{c}}}$ (see Eq.~(\ref{Mc2})), displayed in each plot in Figure~\ref{fig:betalargeT}, denotes the maximum remnant fraction in the dark matter to be one percent (where the rest is just axion dark matter\footnote{\label{footnotehotaxionDM}Note that we also have axion particles emitted through the 2-2-hole radiation. The fraction of these particles in the dark matter energy density, in the domination scenario, is given as $r_a^{2\mbox{-}2}=0.2 B_a \hMmin^{1/2}\hMini^{-1/2}$, where $B_a=g_a/\mathcal{N_*}=1/108$ is the branching fraction of the axion in total number of degrees of freedom~\cite{Aydemir:2020pao}. The largest possible fraction $r_a^{2\mbox{-}2}$ can be found by using the smallest initial 2-2-hole mass in the domination scenario, $\Mini\approx M_c$, and the largest remnant mass, $\Mmin=0.1\;\rm{g}$. This yields $r_a^{2\mbox{-}2}\sim 10^{-13}$, which is not a meaningful contribution and hence can easily be ignored. Note that due to their negligible contribution, these particles are also exempt from the free-streaming constraints~\cite{Aydemir:2020pao}. Yet, these relativistic particles can also be important since they contribute to the effective number of relativistic degrees of freedom, $\Delta N_{\rm{eff}}$, in the early universe, as will be briefly discussed in the last paragraph of this section.}). If we impose even smaller values for this value, then the critical mass $M_{\bar{\textrm{c}}}$ becomes bigger, and less parameter space is available for the axion dark matter in the nonstandard scenario. We also denote the other critical mass $M_{\textrm{c}}$, where the remnants would account for all dark matter, which is not the scenario of interest here. In the case of black holes with no remnants, no such bound exists, i.e. the blue regions disappear; then slightly more parameter space within the red lines in Figure~\ref{fig:beta1} is allowed. This bound, i.e., remnant fraction in the dark matter being negligible, plays more of a role for larger remnant mass for a given axion mass. This is displayed in Figure~\ref{fig:beta2}, where we choose our benchmark value as $\Mmin=0.1\;\rm{g}$ (and the same axion mass as in Figure~\ref{fig:beta1});
 here, there is relatively more overlap between the blue region and the region between red lines. If we impose that the fraction of 2-2-hole remnants in dark matter becomes even lower than 0.01, then $M_{\bar{\textrm{c}}}$ shifts to the right, and at the lowest possible fraction this critical mass equates to $M_{\mathrm{BBN}}$, where there would be no available parameter space. The lowest possible fraction of remnants is obtained for $\Mmin=\Mp$ as approximately $10^{-6}$. As for $\Mmin=0.1\;\rm{g}$. which is the largest allowed remnant mass in the pre-BBN remnant domination scenario, the fraction of the remnants becomes approximately $10^{-5}$.
 
 As illustrated in Figures~\ref{fig:beta3} and \ref{fig:beta4} (as compared to the ones above), for a given $\Mmin$, the smaller the axion mass is the smaller the required parameter space. Note that the required parameter space in Figure~\ref{fig:beta3} consists of a single point; namely the intersection point of $\beta_{\mathrm{max}}$ and $M_{\mathrm{BBN}}$ lines. This can be seen in Figure~\ref{plot1} also, since the axion mass in question, $m_a=3.2\times10^{-9}\;\rm{eV}$, corresponds to the point at which the "knee" occurs, denoting the location where $T_{\rm{osc}}\simeq T_{\rm{QCD}}$.  
 
 In Figure~\ref{fig:betasmallT}, we display the $T_{\rm{osc}}\lesssim T_{\rm{QCD}}$  case, which has a much smaller parameter space than the other case. The only constraint on $\beta$, here, is that it should take values in between the $\beta_{\mathrm{max}}$ and $T_{\rm{osc}}= T_{\rm{eq}}$ lines. The reason for not going below the $T_{\rm{osc}}= T_{\rm{eq}}$ lines is that we cannot have $T_{\rm{osc}}\geqslant T_{\rm{eq}}$ and $T_{\rm{osc}}\lesssim T_{\rm{QCD}}$ together for the largest allowed interval for axion mass, which is below the knee on the red line in Figure~\ref{plot1}; $2.5\times10^{-9}\;\rm{eV} \leqslant  m_a \leqslant 3.2\times 10^{-9}\;\rm{eV}$.\footnote{Recall that the red line in Figure~\ref{plot1} denotes the maximum $\theta_i$ we can get from the hole domination, which corresponds to $\beta=\beta_{\rm{max}}$; for smaller values of $\beta$, the knee gets closer to the $\theta_i=0.5$ line and quickly passes below it. Therefore, the maximum allowed $m_a$ interval for  $T_{\rm{osc}}\lesssim T_{\rm{QCD}}$ is denoted on the red line from the knee to the $\theta_i=0.5$ line.}  This can be seen in the first line of Eq.~(\ref{Tosc1}); for the allowed range of $m_a$, $T_{\rm{osc}}\simeq 0.75 \;\textrm{-}\; 0.85\; \rm{GeV} \geqslant T_{\rm{QCD}}=0.15\;\rm{GeV}$. Since the plot never reaches the $T_{\rm{osc}}= T_{\rm{dom}}$ case, we are always in the $T_{\mathrm{eq}} \geqslant T_{\rm{osc}} \geqslant T_{\mathrm{dom}}$  interval, where there is no $\beta$ dependence in the corresponding equation (given in Eq.~(\ref{axionabundance3})) and hence no constraint on $\beta$ in this region (as denoted by the vertical lines).
 
 We choose two benchmark points to display in Figure~\ref{fig:betasmallT}. In Figure~\ref{fig:beta5}, we again choose $\Mmin=3.6\;\Mp$, which is parametrically the black hole case (modulo the blue region, as explained above). For this figure, we pick the borderline value (namely the knee point in Figure~\ref{plot1}) for the axion mass. Since the allowed region reduces as the axion mass increases, the orange part covers the whole $2.5\times10^{-9}\;\rm{eV} \leqslant  m_a \leqslant 3.2\times 10^{-9}$ region. Therefore, for $m_a=2.5\times10^{-9}\;\rm{eV}$ (for $\Mmin=3.6\;\Mp$) the red line on the left (i.e. $\theta_i=0.5)$ in Figure~\ref{fig:beta5} would reside on $M_{\rm{BBN}}$ line with no allowed (i.e. orange) parameter space left (unless one allows smaller values for the angle). We demonstrate this in Figure~\ref{fig:beta6}. But here, we choose $\Mmin=\Mp$ to be able to go to the lowest $m_a$ allowed (see Eq.~(\ref{mainterval})).
  
Some final remarks are in order. In addition to the cold (DM) axions we have, there are also axions emitted by the 2-2-hole (or black hole) radiation (which do not contribute to dark matter in a meaningful amount, as mentioned in footnote~\ref{footnotehotaxionDM} on the previous page). On the other hand, these nonthermally produced axions are highly relativistic and hence contribute to the effective number of relativistic degrees of freedom, $\Delta N_{\rm{eff}}$. Based on the Planck data~\cite{Planck:2015fie}, the current upper limit is found as $\Delta N_{\mathrm{eff}}\leqslant 0.28$ at 95\% C.L.~\cite{Bernal:2016gxb}. The contribution of particles with a single degree of freedom (hence of the axions) from the 2-2-hole radiation (in the domination scenario) is found as $\Delta N_{\rm{eff}}\simeq 0.02$ (see the purple band in figure 3 of Ref.~\cite{Aydemir:2020pao}), which is consistent with data.
 \\
\\

\section{Summary\label{Conclusion}}

We investigated the effects of a nonstandard cosmological scenario, triggered by the domination of primordial 2-2-holes~\cite{Holdom:2002xy, Holdom:2016nek,Holdom:2019ouz,Ren:2019afg} in the early universe, on axion dark matter. In comparison to the PBH counterpart~\cite{Bernal:2021yyb}, we have an extra parameter, namely the remnant mass $\Mmin$, which is directly related to the underlying quantum gravity, quadratic gravity~\cite{Stelle:1976gc, Voronov:1984kq, Fradkin:1981iu, Avramidi:1985ki}.  Most importantly, the remnant mass appears also in the classical Hawking-like radiation of  2-2-holes, which lasts until they go on the remnant stage. These remnants themselves are viable dark matter candidates, as we studied before~\cite{Aydemir:2020xfd}. However, in this paper, we focused on "the axion as (almost) all dark matter scenario", and thus we ensured through our parameter selection that the 2-2-hole remnant contribution to dark matter was negligible.

 Due to the modification in the evolution of the universe as a result of corresponding changes in the Hubble parameter in different temperature intervals, the abundance constraints of the axion dark matter changes. We found that the lower limit on the axion mass becomes as low as $m_a \sim10^{-9}$ eV (as opposed to the standard scenario value of $m_a \sim10^{-6}$ eV) for the Planck mass remnants, which is the case for a strongly coupled quantum gravity. Furthermore, the domination scenario itself constrains the remnant mass $\Mmin$, considerably. Given that we focused on the pre-BBN domination scenario in order not to interfere with BBN (Big Bang nucleosynthesis) constraints, the remnant mass window becomes $\Mp \lesssim \Mmin \lesssim 0.1\;\rm{g}$.  We also discussed the implications of this scenario on the initial fraction of holes ($\beta$) in energy density, where we took into account the corresponding gravitational wave constraints~\cite{Domenech:2020ssp}.
 
Finally, we note that the study of such black-hole mimickers constitutes an example of the effects of these BH alternatives on any related phenomenon. Since these objects appear in theories beyond General Relativity, any sign of their existence may have profound effects on our current knowledge of gravity. The fact that the remnant mass $\Mmin$ is a parameter directly related to the underlying theory gives a chance to constrain quantum gravity. Therefore, it is important to study available candidates and the corresponding phenomenological implications. 
\section*{Acknowledgment}
The work of U.A. is funded by The Scientific and Technological Research Council of T\"urkiye (T\"UB\.ITAK) B\.{I}DEB 2232-A program under project number 121C067. \\
\\

\appendix

\section{Brief review on 2-2-holes\label{22holereview}}
Here, we briefly review 2-2-holes. More detailed discussions can be found in Refs.~\cite{Holdom:2016nek,Holdom:2019ouz,Ren:2019afg,Aydemir:2020ezy}. The action of quadratic gravity is given as
\begin{eqnarray}
\label{action}
S_{\mathrm{QG}}= \frac{1}{16\pi}\int d^4 x \sqrt{-g}\left(m_{\mathrm{Pl}}^2 R-\alpha\; C_{\mu\nu\rho\sigma}C^{\mu\nu\rho\sigma}+\beta R^2\right)\,,
\end{eqnarray}
where $\alpha$ and $\beta$ are dimensionless couplings. The new terms, namely the Ricci scalar square and the Weyl tensor square,  bring in, in addition to the usual massless graviton, a new spin-0 and a spin-2 mode with the tree level masses $m_0\approx \Mp/\sqrt{\beta}$ and $m_2\approx \,\Mp/\sqrt{\alpha}$, respectively. The quadratic theory is renormalizable and asymptotically free~\cite{Stelle:1976gc, Voronov:1984kq, Fradkin:1981iu, Avramidi:1985ki}  and can also be obtained from the string theory as an effective field theory~\cite{Alvarez-Gaume:2015rwa}. Yet, the theory suffers from an Ostrogradsky instability, associated with a  spin-2 ghost. The proposed solutions, in general, employ modifications to quantum prescription, depending on whether the theory becomes strongly or weakly coupled at the Planck scale~\cite{Tomboulis:1977jk, Grinstein:2008bg, Anselmi:2017yux, Donoghue:2018lmc, Bender:2007wu, Salvio:2015gsi, Holdom:2015kbf, Holdom:2016xfn, Salvio:2018crh, Alvarez-Gaume:2015rwa, Donoghue:2021cza, Donoghue:2021eto}. 

The theory generally admits spherically symmetric solutions where the behavior around the origin is in the form $g_{tt}=b_m  r^m+..$ and  $g_{rr}=a_n r^n+..$, which are characterized as $(n,m)$. In addition to the expected family of solutions that have analogs in GR such as the black hole $(1,-1)$ and star solutions $(0,0)$, the quadratic theory also has the $(2,2)$ solutions, hence the name 2-2-holes~\cite{Holdom:2016nek}. These objects can be as compact as black holes but without an event horizon. From the exterior, they closely resemble black holes, whereas in the interior, the 2-2-hole solution takes over.
The importance is that in GR, such ultracompact configurations are not supported, and the endpoint of the gravitational collapse is a black hole, which is, consequently, the most compact object in GR; in quadratic gravity, on the other hand, 2-2-hole solutions are also a candidate for the endpoint. 

The term responsible for the 2-2-hole solutions is the Weyl term $C^{\mu\nu\rho\sigma}C_{\mu\nu\rho\sigma}$ in the quadratic action, given in Eq.~(\ref{action}), whereas the $R^2$ term is optional and therefore could be neglected for simplicity~\cite{Holdom:2016nek,Holdom:2019ouz,Ren:2019afg}. Thermal 2-2-holes, sourced by a thermal gas, can be arbitrarily large, but there is no solution below the value $\approx \frac{\Mp^2}{ m_2}$~\cite{Holdom:2019ouz,Ren:2019afg}. Consequently, one can state that the new spin-2 mode, generated by the Weyl term, determines the minimum allowed mass for the 2-2-hole. We define the parameter $\hMmin$, to be used throughout the paper, as

\begin{eqnarray}
\label{eq:Mmin}
\hMmin\equiv\frac{\Mmin}{\Mp}\approx \frac{\Mp}{ m_2}\approx  \frac{\lambda_2}{\lp};
\end{eqnarray}
where $\lambda_2$ is the corresponding Compton wavelength. There exist two main scenarios for quadratic gravity regarding the strength of dimensionless couplings in action~(\ref{action}). In the strong coupling scenario, the Planck mass is expected to emerge dynamically through dimensional transmutation as the only mass scale, $m_2\approx \Mp$, i.e. $\hMmin\approx 1$. On the other hand, in the weak coupling scenario, it can be generated spontaneously through vacuum expectation values of some scalar fields or introduced explicitly~\cite{Holdom:2019ouz}. In this case, there can be a large mass hierarchy with $m_2\ll \Mp$, i.e. $\hMmin\gg 1$. 

To investigate the thermodynamics of such solutions, one can focus on massless  particles, with the equation of state
\begin{equation}
\rho=3p=\frac{\pi^2}{30} \mathcal{N}\, T^{4}\,,
\end{equation}
where $\rho$ and $p$ are the energy density and pressure, respectively. $T(r)$ denotes the local temperature and 
$\mathcal{N}= g_b+7 g_f/8$, where $g_b$ and  $g_f$ are the numbers of bosonic and fermionic degrees of freedom.
Since in the interior, $T(r)$ reaches arbitrarily high values, $\mathcal{N}$ accounts for particle species of any mass and, in principle, could be much larger than its Standard Model value $\mathcal{N}\approx 107$ if there are new particles in Nature. 
The conservation law of the stress tensor, as usual, leads to Tolman law ($T(r)g_{tt}^{1/2}=T_\infty$), where the value at spatial infinity, $T_\infty$, is roughly the temperature measured by a distant observer. 

The temperature for a normal (nonremnant) 2-2-hole is determined to be~\cite{Ren:2019afg}  
\begin{eqnarray}
\label{22holetemp} T\simeq 1.7\, \mathcal{N}^{-1/4}\hMmin^{1/2}\, T_\textrm{BH}\equiv\mathcal{A} \;T_\textrm{BH},\end{eqnarray}
where $\hMmin\equiv \Mmin/\Mp$ and the Hawking temperature $T_\textrm{BH}=\Mp^2/8\pi M$. Here, we introduced the symbol $\mathcal{A}$ to use in the main text.  The entropy is given as
\begin{eqnarray}
\label{22holeentropy} S&\simeq& 0.60\, \mathcal{N}^{1/4}\hMmin^{-1/2} \,S_\textrm{BH}\,,
\end{eqnarray}
where the Bekenstein--Hawking entropy  $S_\textrm{BH}=\pi\, r_H^2/\lp^2$. At this stage, the thermodynamic behavior of a 2-2-hole significantly resembles black hole thermodynamics; it exhibits Hawking-like radiation with a negative heat capacity and fulfills the area law for entropy. Remarkably, this behavior of the thermal 2-2-hole directly arises from self-gravitating relativistic thermal gas on a highly curved background spacetime without considering spontaneous particle creation from the vacuum and, therefore, originates from a completely different origin than the black holes.

The above relations are for the nonremnant stage of the 2-2-hole, where the hole's mass is large (i.e. away from the minimum mass). Once the temperature reaches the peak value, which happens at around $M\simeq 1.2 \Mmin$~\cite{Ren:2019afg}, the 2-2-hole enters
into the remnant stage, where drastic changes occur in the
thermodynamic behavior; the object starts to radiate like a regular object with heat capacitypositive. The evaporation continues extremely slowly and asymptotically halts. 

The mass evolution of a thermal 2-2-hole can be described by the Stefan--Boltzmann law
\begin{equation}\label{eq:SBlaw}
-\frac{dM}{dt}
\simeq \frac{\pi^2}{120}\, \mathcal{N}_* \, 4\pi r_H^2 \,T^4\;,
\end{equation}
where $4\pi r_H^2$ is the effective emitted area. $\mathcal{N}_*$ accounts for the number of particles lighter than $T$~\cite{MacGibbon:1991tj}, and it could be much smaller than $\mathcal{N}$. 
The time dependences of the temperature and mass take the same form as for a black hole. Treating $\mathcal{N}_*$ as a constant determined by the initial $T$, we have in the nonremnant stage that 
\begin{eqnarray}\label{eq:LMlimitTime0}
T_{\infty}(t)\simeq T_{\infty,\textrm{init}}\left(1-\frac{\Delta t}{\tau_L}\right)^{-1/3},\quad
M(t)\simeq \Mini\left(1-\frac{\Delta t}{\tau_L}\right)^{1/3},
\end{eqnarray}
where $\tau_L$ is the evaporation time for a 2-2-hole evolving from a much larger $\Mini$ to $\Mmin$,
\begin{eqnarray}\label{eq:tauL}
\tau_L
\,\simeq \,2\times 10^{-40} \, \mathcal{N}\mathcal{N_*}^{-1}  \,  \hMmin^{-2} \, \hMini^{3}\; \textrm{s}\,,
\end{eqnarray}
where $\hMmin\equiv \Mmin/\Mp$ and $\hMini\equiv \Mini/\Mp$.
\bibliographystyle{JHEP}

\providecommand{\href}[2]{#2}\begingroup\raggedright\endgroup

\end{document}